\documentclass[namedreferences]{solarphysics}
\usepackage[hyperref,optionalrh,solaromanenum]{spr-sola-addons} % For Solar Physics 
\usepackage{graphicx}                    % For eps figures, newer & more powerfull
\usepackage{amssymb}                    % useful mathematical symbols
\usepackage{color}                       % For color text: \color command
\usepackage{breakurl}                         % For breaking URLs easily trough lines
\begin{document}

\begin{article}

\begin{opening}

\title{How good is the bipolar approximation of active regions for surface flux transport?}

%%%%%%%%%%%%%%%%%%%%%%%%%%%%%%%%%%%%%%%%%%%%%%%%%%%
%% Authors Names
%
% \author[addressref={},corref,email={}]{\inits{}\fnm{}\lnm{}\orcid{}}
\author[email={anthony.yeates@durham.ac.uk}]{\inits{A.R.}\fnm{Anthony R.}\lnm{Yeates}\orcid{0000-0002-2728-4053}}

%%%%%%%%%%%%%%%%%%%%%%%%%%%%%%%%%%%%%%%%%%%%%%%%%%%
%% Runningheads
%
%\runningauthor{}
%\runningtitle{}

%%%%%%%%%%%%%%%%%%%%%%%%%%%%%%%%%%%%%%%%%%%%%%%%%%%
%% Affilations 
%% id shold be the same with \author addressref value.
\address{Department of Mathematical Sciences, Durham University, Durham, DH1 3LE, UK}

%%%%%%%%%%%%%%%%%%%%%%%%%%%%%%%%%%%%%%%%%%%%%%%%%%%
%%% Abstract 
\begin{abstract}

We investigate how representing active regions with bipolar magnetic regions (BMRs) affects the end-of-cycle polar field predicted by the surface flux transport model.
Our study is based on a new database of BMRs derived from the SDO/HMI active region patch data between 2010 and 2020.
An automated code is developed for fitting each active region patch with a BMR, matching both the magnetic flux and axial dipole moment of the region and removing repeat observations of the same region.
By comparing the predicted evolution of each of the 1090 BMRs with the predicted evolution of their original active region patches, we show that the bipolar approximation leads to a 24\% overestimate of the net axial dipole moment, given the same flow parameters.
This is caused by neglecting the more complex multipolar and/or asymmetric magnetic structures of many of the real active regions, and may explain why previous flux transport models had to reduce BMR tilt angles to obtain realistic polar fields.
Our BMR database and the Python code to extract it are freely available.
\end{abstract}

%%%%%%%%%%%%%%%%%%%%%%%%%%%%%%%%%%%%%%%%%%%%%%%%%%%
%% Keywords
%
%\keywords{}

\end{opening}
%-------------------------------------------------

%%%%%%%%%%%%%%%%%%%%%%%%%%%%%%%%%%%%%%%%%%%%%%%%%%%
%% Sections
%
\section{Introduction}\label{s:intro} 

Despite its simplicity, the surface flux transport (SFT) model introduced by \citet{leighton1964} has proved remarkably effective at mimicking the evolution of the large-scale magnetic field on the solar surface \citep[see the reviews by][]{sheeley2005,jiang2014,wang2017}. An important recent application is solar cycle prediction. Since the polar field at Solar Minimum is the best predictor for the following solar cycle amplitude \citep{schatten1978,munoz2013,pesnell2016,petrovay2020}, the SFT model offers the possibility of extending predictions earlier since it -- in turn -- predicts this polar field. Several predictions of Cycle 25 have been made with this methodology \citep{iijima2017,jiang2018,upton2018}, as well as by coupling SFT to interior dynamo models \citep{bhowmik2018,labonville2019}.

A major component of randomness in the solar cycle is now believed to arise from active region emergence. There are significant fluctuations both in the number of emerging active regions and in their properties such as emergence latitude, magnetic flux, or tilt angle \citep[see the review by][]{vandriel2015}. General cycle-to-cycle trends remain unclear, with some studies suggesting active regions in stronger solar cycles to have lower tilt angles \citep[\textit{e.g.},][]{mcclintock2013} and others finding no significant variation \citep{tlatova2018}. Such a trend would act to stabilise the solar cycle by reducing the polar field strength at the end of strong cycles. Another stabilising effect is the tendency for active regions in stronger cycles to emerge at higher latitudes, reducing the cross-equatorial flux transport needed to build the polar field \citep{jiang2014}.  On the other hand, it is now believed that individual ``rogue'' active regions with statistically extreme properties can have a significant effect on the polar field \citep{cameron2013,jiang2014a,nagy2017, whitbread2018}.

To account for these fluctuations within particular solar cycles, SFT models need to be driven with input data tailored to real observations of the Sun, rather than generic statistical distributions. The most direct method is to assimilate observed magnetograms directly from observed regions of the Sun \citep[\textit{e.g.,}][]{worden2000,upton2014,hickman2015}. However, for many applications it is useful to isolate individual active region sources. Reasons to do this might include: efficiency and speed of calculation, particularly when running ensembles for prediction purposes \citep{petrovay2020a}; coupling to three-dimensional models for either the solar corona \citep{mackay2012,yeates2014} or interior \citep{bhowmik2018,labonville2019}; determining the contribution of individual active regions \citep{yeates2015,whitbread2018}; and extending simulations to previous epochs where sunspot observations are available but magnetograms are not \citep{virtanen2019p4}. Whilst some authors have adopted a compromise approach of assimilating magnetograms for individual regions \citep{yeates2015,whitbread2017,virtanen2019p4}, the majority of models fit each source region with a simple bipolar magnetic region (BMR). Public catalogues of magnetic BMRs determined from NSO data are available  for Cycle 21 from \citet{wang1989} and Cycles 23-24 from \citet{yeatesdata}.

This paper concerns the consequences fitting active regions with (symmetric) BMR shapes, when these BMRs are tailored to individual observed active regions.
A difficulty with this approach has been found when the magnetic flux, size and tilt angle of each BMRs are all matched to the corresponding observed active region. Using the observed values for all of these properties tends to lead to an over-strong polar field after the ensuing SFT evolution \citep[\textit{e.g.},][]{yeates2014,cameron2010,jiang2014a,bhowmik2018}. To correct the discrepancy, these authors reduce the tilt angles of the inserted BMRs compared to their observed values, with a typical reduction of 20\% to 30\%.
This reduction is argued to compensate for the lack of inflows towards active regions in the SFT model, which could have the effect of reducing the tilt angle during subsequent evolution of the real region \citep{cameron2010}.
The BMR-driven SFT model for Cycle 24 described in this paper also overestimates polar field. However, by comparing with the same model driven by non-BMR sources, we will show that this overestimate results simply from the assumption of bipolar shape, rather than any systematic error in the transport \textit{per se}.

A hint of the limitations of bipolar shape comes from the recent work of \citet{iijima2019} on the asymmetry of bipolar regions. These authors explore the consequences of the fact that, in real active regions, leading polarity flux tends to be more spatially concentrated than following polarity flux \citep{fisher2000}. \citet{iijima2019} used an SFT model to show that if the following polarity is more dispersed than the leading polarity, then sufficient following flux can escape into the opposite hemisphere to reduce the predicted polar field. Here, we test what is lost in the BMR approximation more generally, using a new automated BMR database for Cycle 24 (Section \ref{s:database}). The effect of fitting BMRs on the SFT evolution is discussed in Section \ref{s:predicted}, and concluding remarks in Section \ref{s:conc}.

\section{Bipole database}\label{s:database}

Our database is derived from the Spaceweather HMI Active Region Patch (SHARP) data from the Helioseismic and Magnetic Imager on Solar Dynamics Observatory \citep{Bobra2014}. Each SHARP represents an individual active region tracked over a single disk passage. Using SHARP data, we avoid the need to repeat the identification of individual active regions (except for regions with more than one disk passage), and can take advantage of the existing metadata to query the vast HMI database. We use the \texttt{hmi.sharp\_cea\_720s} series, which comprise definitive data remapped to a cylindrical equal-area projection. Here, ``definitive'' means that the regions are identified and defined only after their full disk passage. This series is chosen because it includes vector magnetic field data (and derived quantities) that may be included in our database in future -- for example, to estimate the helicity of BMRs. However, for the initial database described in this paper, we use only the (remapped) line-of-sight magnetic field.

Our open-source Python code for BMR extraction is available online at \url{https://github.com/antyeates1983/sharps-bmrs}, and the resulting database generated for this paper is available on the Harvard Dataverse \citep{yeatesdata2020}. We stress that the philosophy in developing this new BMR database has been to minimise subjectivity through development of a fully-automated BMR fitting code.

\subsection{Magnetogram extraction}

\begin{figure}
\includegraphics[width=\textwidth]{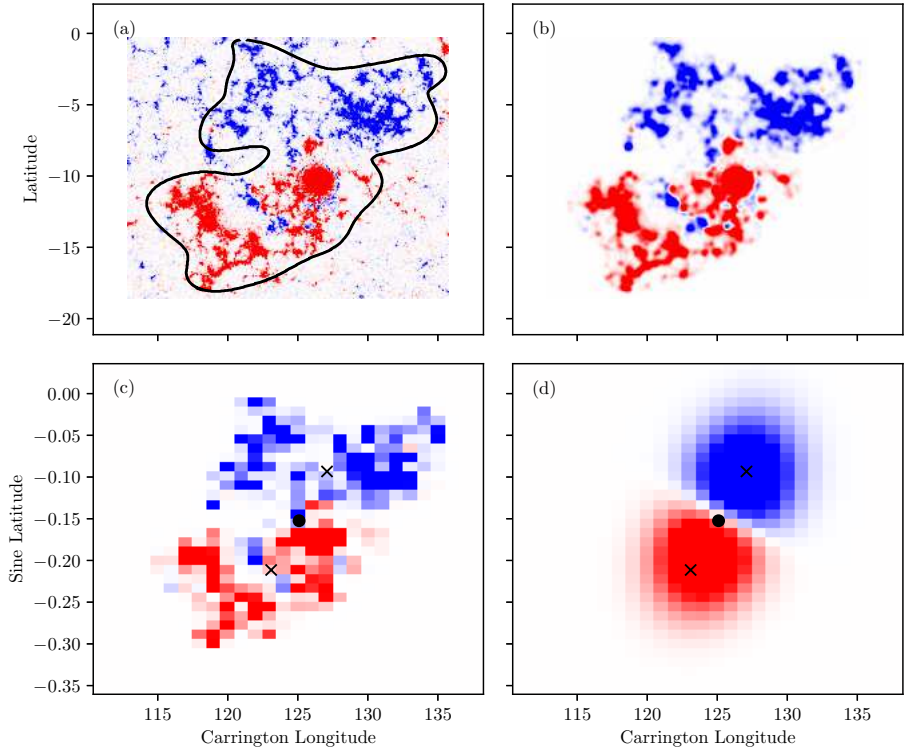}
\caption{An example SHARP region, number 7147 which passed Central Meridian on 30 September 2017. Panel (a) shows the HMI line-of-sight magnetogram with the provided SHARP mask indicated in black. Panel (b) shows the result of the Gaussian smoothing. Panel (c) shows the interpolated magnetogram on the computational grid, and panel (d) shows the approximating BMR. All panels are saturated at $\pm 100\,\mathrm{G}$ (red positive, blue negative). This region is in the Southern hemisphere, so is an ``anti-Hale'' region for Cycle 24. In (c) and (d), the overall centroid $(s_0,\phi_0)$ is shown by the black circle and the polarity centroids $(s_+,\phi_+)$, $(s_-,\phi_-)$ by the black crosses.}
\label{fig:harp}
\end{figure}

For each SHARP, we select a single representative magnetogram for our database. Each SHARP has multiple associated magnetograms, observed at high cadence over the region's disk passage. Because we are using line-of-sight magnetic data, we select the observation with flux-weighted centroid closest to Central Meridian. This information is conveniently available in the SHARP metadata. One such magnetogram is illustrated in Figure \ref{fig:harp}(a).

Having downloaded a single line-of-sight magnetogram for the given SHARP, we set pixels outside the masked region to zero. Then, we apply a Gaussian smoothing and interpolate the data to a lower resolution grid that is more appropriate for global simulations. The results in this paper, and the published database, use a uniform grid of $180\times 360$ cells in sine-latitude and longitude, although this resolution can be modified in the accompanying code, as can the parameters of the Gaussian filter. We took $\sigma=4$ pixels for the standard deviation of the Gaussian kernel, and used cubic interpolation. The results of these two stages are illustrated in Figure \ref{fig:harp}(b) and (c).

Once interpolated to the computational grid, we compute and record the flux imbalance, defined as the ratio of net flux to absolute flux over all pixels $i,j$,
\begin{equation}
\Delta\Phi = \frac{\sum_{i,j}B^{i,j}}{\sum_{i,j}|B^{i,j}|}.
\end{equation}
Note that all cells have equal area on this grid. The original SHARP regions are not flux balanced, so, once interpolated, regions will lie somewhere between $\Delta\Phi=0$ (perfectly flux balanced) and $\Delta\Phi=1$ (unipolar). Regions with large imbalance will be discarded from the resulting BMR database, as discussed in Section \ref{s:filter}. For a region that is retained, the magnetogram is corrected for flux balance by applying different multiplicative scaling factors to the positive and negative pixels, in such a way that both the positive and negative fluxes are scaled to the original mean of the two, and the overall unsigned flux is unchanged. This ensures that the polarity inversion lines do not change position. 

\subsection{Bipolar approximation}

To compute the approximating bipolar magnetic region (BMR) for a given SHARP, we first compute the centroids $(s_+, \phi_+)$ and $(s_-,\phi_-)$ of positive and negative $B$ on the computational grid. Here $s$ denotes sine-latitude and $\phi$ denotes (Carrington) longitude. Based on these polarity centroids, we compute (i) the overall centroids
\begin{equation}
s_0 = \frac12(s_+ + s_-),\qquad \phi_0 = \frac12(\phi_+ + \phi_-),
\end{equation}
(ii) the polarity separation, which is the heliographic angle
\begin{equation}
\rho = \arccos\left[s_+s_- + \sqrt{1-s_+^2}\sqrt{1 - s_-^2}\cos(\phi_+-\phi_-) \right],
\end{equation}
and (iii) the tilt angle with respect to the equator, given by
\begin{equation}
\gamma = \arctan\left[\frac{\arcsin(s_+) - \arcsin(s_-)}{\sqrt{1-s_0^2}(\phi_- - \phi_+)}\right].
\label{eqn:tilt}
\end{equation}
Together with the unsigned flux, $|\Phi|$, these parameters define the BMR for our chosen functional form. For an untilted BMR centered at $s=\phi=0$, this functional form is defined as
\begin{equation}
B(s,\phi) = F(s,\phi) = -B_0\frac{\phi}{\rho}\exp\left[-\frac{\phi^2 + 2\arcsin^2(s)}{(a\rho)^2}\right],
\label{eqn:bmr}
\end{equation}
where the amplitude $B_0$ is scaled to match the corrected flux of the observed region on the computational grid. To account for the location $(s_0,\phi_0)$ and tilt $\gamma$ of a general region, we set $B(s,\phi) = F(s',\phi')$, where $(s',\phi')$ are spherical coordinates in a frame where the region is centered at $s'=\phi'=0$ and untilted (explicit expressions are given in Appendix \ref{a:rotation}). Figure \ref{fig:harp}(d) shows an example BMR.

The parameter $a$ in Equation (\ref{eqn:bmr}) controls the size of the BMR relative to the separation, $\rho$, of the original polarity centroids. For given values of $\lambda_0$, $\gamma$, and $\rho$, and $B_0$ chosen to give the required magnetic flux, the parameter $a$ may be chosen to control the axial dipole moment of the BMR. In Section \ref{s:properties}, we will see that a good match to the axial dipole moment of the original SHARP is obtained with $a=0.56$, and the same value works for every region.

\subsection{Filtering} \label{s:filter}

To ensure that only meaningful BMRs are included in the database, we filter out those SHARPs that do not meet the following criteria:
\begin{enumerate}
\item The flux imbalance (before correction) should be less than a given threshold -- we choose $\Delta\Phi \leq 0.5$. This effectively discounts unipolar SHARPs, which cannot be approximated by BMRs.
\item The polarity separation should be resolved on the computational grid, \textit{i.e.}, $\rho \ge \Delta\phi$ where $\Delta\phi$ is the longitudinal grid spacing.
\end{enumerate}
The third and fourth columns of Table \ref{tab:filter} show the number of SHARPs per year which were rejected at each of these two filtering steps. The figures in parentheses indicate total unsigned flux in Mx. Overall, about 13\% of the SHARP flux is rejected as being unipolar (step i), and only a further 0.3\% because of insufficient polarity separation (step ii). Following these two steps, we further identify and remove regions that correspond to repeat observations. These are shown in the fifth column of Table \ref{tab:filter}, and our procedure for identifying them is described next.

\begin{table}
\caption{Number of identified and filtered regions (parentheses show  unsigned flux in Mx).}
\label{tab:filter}
\begin{tabular}{lrrrrr}
\hline
Year & SHARPs & \multicolumn{3}{c}{Rejected SHARPs} & BMRs\\
 &  & Unipolar & Unseparated & Repeated & \\
\hline
2010* & 170 (3.7e23) & 103 (7.8e22) & 8 (1.1e21) & 4 (4.4e22) & 55 (2.5e23)\\
2011 & 458 (1.9e24) & 255 (1.7e23) & 18 (2.7e21) & 25 (2.3e23) & 160 (1.5e24)\\
2012 & 548 (2.1e24) & 357 (2.9e23) & 13 (1.9e21) & 22 (1.9e23) & 156 (1.6e24)\\
2013 & 617 (2.3e24) & 380 (2.7e23) & 18 (3.9e21) & 20 (1.8e23) & 198 (1.9e24)\\
2014 & 637 (3.3e24) & 416 (3.7e23) & 17 (1.8e22) & 23 (4.3e23) & 181 (2.5e24)\\
2015 & 599 (2.2e24) & 425 (3.8e23) & 12 (3.7e21) & 29 (4.3e23) & 133 (1.4e24)\\
2016 & 357 (9.1e23) & 226 (1.6e23) & 16 (4.8e21) & 12 (8.9e22) & 103 (6.6e23)\\
2017 & 163 (4.8e23) & 99 (5.9e22) & 6 (1.1e20) & 7 (5.2e22) & 51 (3.6e23)\\
2018 & 74 (1.2e23) & 41 (2.1e22) & 4 (8.6e20) & 0 (0.0e00) & 29 (1.0e23)\\
2019 & 41 (8.3e22) & 19 (1.8e22) & 2 (3.9e20) & 1 (4.8e21) & 19 (5.9e22)\\
2020* & 7 (1.0e22) & 2 (1.8e20) & 0 (0.0e00) & 0 (0.0e00) & 5 (1.0e22)\\
[0.2cm]
Total: & 3671 (1.4e25) & 2323 (1.8e24) & 114 (3.8e22) & 143 (1.7e24) & 1090 (1.0e25)\\
\hline
\end{tabular}\\[0.2cm]
* Coverage of the years 2010 and 2020 is incomplete: the first SHARP included is on 3 May 2010 and the last on 5 April 2020.
\end{table}

\subsection{Removal of repeat observations} \label{s:repeats}

\begin{figure}
\includegraphics[width=\textwidth]{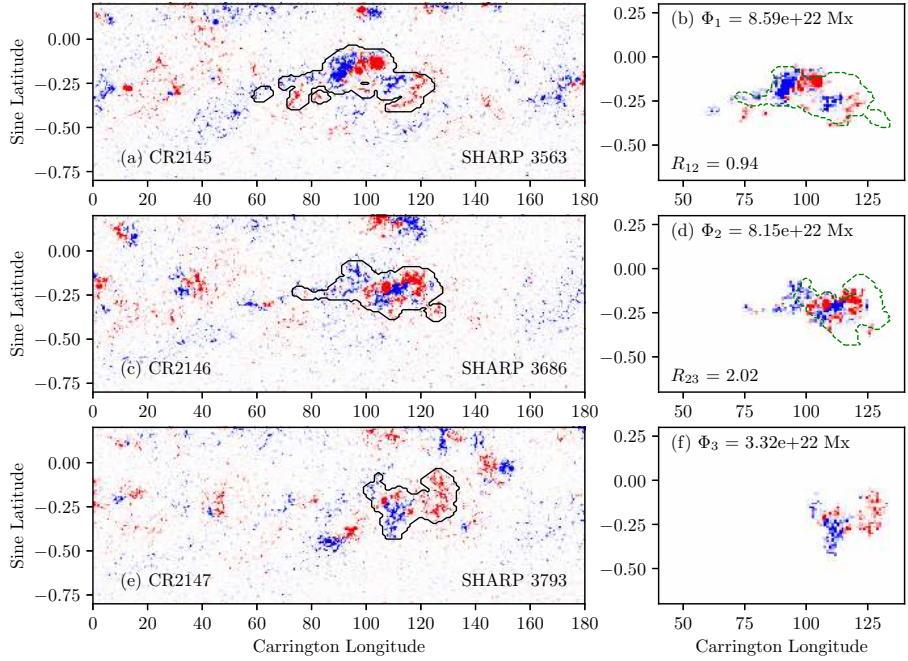}
\caption{Two successive pairs of candidate repeat regions: SHARPs 3563/3686 and  3686/3793. The left column shows the regions in context overlaid on  HMI synoptic magnetograms (with the SHARP masks in black). The right column shows each SHARP on the computational grid. Dashed green contours in (b) and (d) show the derotated boundaries of the later SHARPs, 3686 and 3793 respectively. In all cases radial magnetic field $B$ is shown with red positive and blue negative, saturated at $\pm 200\,\mathrm{G}$. The fluxes $\Phi_1$, $\Phi_2$, $\Phi_3$ give the total unsigned flux for each SHARP.
}
\label{fig:olap}
\end{figure}

To avoid double-counting BMRs in our database, we have implemented a procedure to remove SHARPs that correspond to a repeat observation of a decaying region already in the database from a previous disk passage. This is done by comparing each SHARP with every existing SHARP that passed Central Meridian between 20 and 34 days earlier. To illustrate the procedure, Figure \ref{fig:olap} shows a chain of three SHARPs where both successive pairs are candidate repeat regions.

To determine automatically whether a SHARP $B^{i,j}_n$ is a repeat observation of an earlier region $B^{i,j}_m$, we first derotate $B^{i,j}_n$ to remove the effect of differential rotation, producing a SHARP $\overline{B}^{i,j}_n$ that may be directly compared with $B^{i,j}_m$. The derotation uses the (Carrington frame) angular velocity profile
\begin{equation}
\Omega(s) = \Omega_A + \Omega_Bs^2 + \Omega_Cs^4,
\end{equation}
where $\Omega_A=0.18^\circ\,\mathrm{day}^{-1}$, $\Omega_B=-2.396^\circ\,\mathrm{day}^{-1}$, $\Omega_C=-1.787^\circ\,\mathrm{day}^{-1}$. We then classify $B^{i,j}_n$ as a repeat observation of $B^{i,j}_m$ if $B^{i,j}_m$ had higher unsigned flux than $\overline{B}^{i,j}_n$ in the latter's envelope region $\overline{D}_n$. Specifically, if $R_{nm}>1$ where 
\begin{equation}
R_{nm} = \frac{\sum_{\overline{D}_n}|B^{i,j}_m|}{\sum_{\overline{D}_n} |\overline{B}^{i,j}_n|}.
\end{equation}
For the two successive pairs in Figure \ref{fig:olap}, we find $R_{12}=0.94$ and $R_{23}=2.02$. Although most of the flux of SHARP 3563 lies within the derotated envelope of SHARP 3686 (dashed green line in Figure \ref{fig:olap}b), the ratio $R_{12}$ is less than unity because there is still rather strong flux present in SHARP 3686, and there is a sufficient mismatch in its (derotated) position compared to the earlier region. Thus SHARP 3686 is classified as a new region. On the other hand, $R_{23}$ is substantially greater than unity, so SHARP 3793 is classified as a repeat observation of SHARP 3686. Both decisions concur with a manual determination from the original HMI magnetograms in the left column. 

The numbers of repeat regions identified in each year are shown in Table \ref{tab:filter}. There are 143 repeat regions identified in total, with about 12\% of the original SHARP flux removed in this final stage of the filtering procedure. This leaves 1090 BMRs in the database, with a total unsigned flux of $1.0\times 10^{25}\,\mathrm{Mx}$.

\subsection{Summary of observed BMR properties} \label{s:properties}

\begin{figure}
\includegraphics[width=\textwidth]{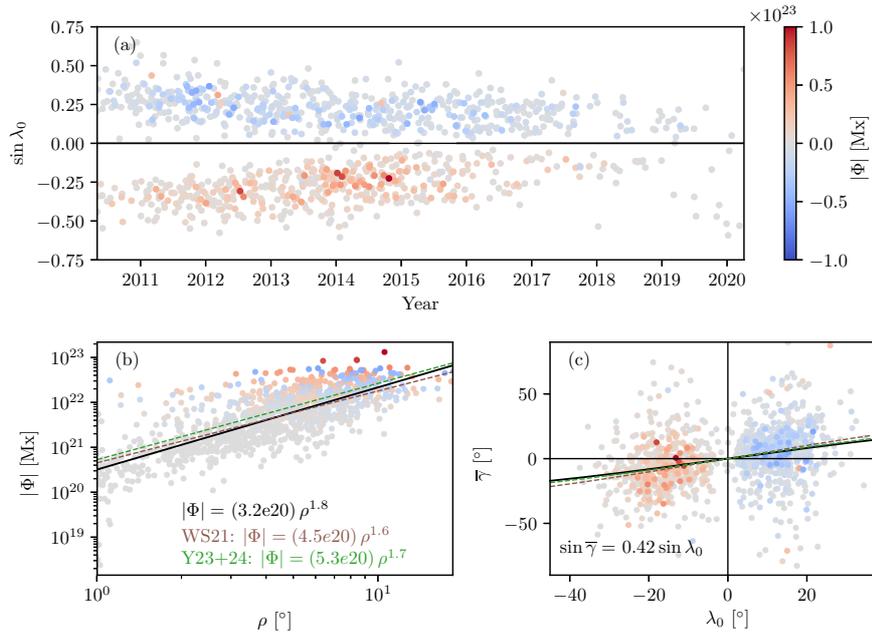}
\caption{Properties of the 1090 BMRs in the database, including (a) sine-latitude versus time; (b) unsigned flux versus polarity separation; and (c) tilt angle against latitude. In all three plots, points are coloured by unsigned flux, with the sign indicating the leading polarity. Panel (c) shows the unsigned tilt angle $\overline{\gamma}$. Black curves in (b) and (c) show least-squares fits as indicated (over all BMRs), while dashed curves show similar fits to the BMR databases of \citet{wang1989} for Cycle 21 and \citet{yeatesdata} for Cycles 23 and 24. In (c) the fit coefficients are  $0.52$ for WS21 and $0.45$ for Y23+24.
}
\label{fig:bmr_params}
\end{figure}

Figure \ref{fig:bmr_params} summarizes the main properties of the 1090  BMRs in the database. In all three panels, the blue/red colour indicates the sign of the leading (Westward) polarity, so that Hale's law is evident in Figure \ref{fig:bmr_params}. For comparison with previous studies, Figures \ref{fig:bmr_params}(b) and (c) show flux versus polarity separation, and tilt versus latitude, respectively. Note that the tilt angle shown here has been adjusted to lie within the range $\pm 90^\circ$ by disregarding the magnetic polarity. Thus we plot the unsigned tilt angle
\begin{equation}
\overline{\gamma} = \left\{
\begin{array}{ll}
\gamma + 180^\circ & \textrm{for $\gamma < -90^\circ$},\\
\gamma & \textrm{for $-90^\circ \leq \gamma\leq 90^\circ$},\\
\gamma - 180^\circ & \textrm{for $\gamma > 90^\circ$}.
\end{array}
\right.
\end{equation}

The least-squares fit in Figure \ref{fig:bmr_params}(b) shows that the unsigned flux scales as $\rho^{1.8}$. Here $|\Phi|$ is the total absolute flux over both polarities. This is consistent with flux being roughly proportional to the area of an active region. The dashed lines show fits for two other BMR catalogues: (i) for Cycle 21, derived from Kitt Peak Vacuum Telescope (KPVT) full-disk magnetograms \citep{wang1989, wangdata}, and (ii) for Cycles 23 and 24, derived from KPVT and later SOLIS/VSM synoptic magnetograms \citep{yeatesdata}. The steeper power law in our new database seems to arise from the scatter in flux at the small-$\rho$ end, which may be affected by our resolution and the thresholds applied to define SHARPs. 
As shown by Figure \ref{fig:bmr_params}(c), our least-squares fit for tilt angle (Joy's law) is close to that in the \citet{yeatesdata} database. The slope found for the \citet{wang1989} BMRs is a little higher, but given the spread in tilt angles, it is difficult to be sure of the statistical significance of this difference. The fits are also broadly in line with \citet{stenflo2012}, who found $\overline{\gamma} = 32.1^\circ\sin\lambda_0$ for Cycle 23 using MDI data. 

\begin{figure}
\includegraphics[width=\textwidth]{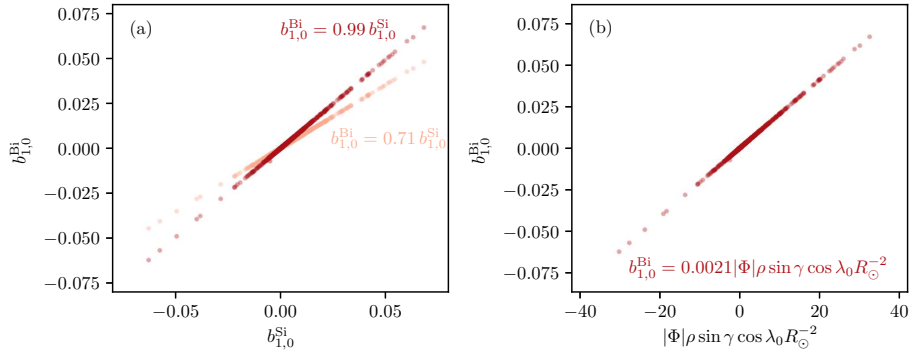}
\caption{Scatter plots of BMR (initial) axial dipole moment against (a) the (initial) axial dipole moment of the original SHARP region, and (b) the parameter combination $|\Phi|\rho\sin\gamma\cos\lambda_0R_\odot^{-2}$. Least-squares linear fits are indicated. In (a), the fainter points show the effect of reducing the BMR size in Equation (\ref{eqn:bmr}) from $a=0.56$ to $a=0.4$.
}
\label{fig:dipi}
\end{figure}

For our purposes in Section \ref{s:predicted}, an important BMR parameter is the axial dipole moment,
\begin{equation}
b_{1,0} =  \frac{3}{4\pi}\int_0^{2\pi}\int_{-1}^1 sB(s,\phi)\,\mathrm{d}s\,\mathrm{d}\phi.
\label{eqn:dip}
\end{equation}
For each region, we record both $b_{1,0}^{\rm Bi}$ -- the axial dipole moment of the fitted BMR -- and $b_{1,0}^{\rm Si}$ -- the axial dipole moment of the original SHARP region on the computational grid (as in Figure \ref{fig:harp}c). 
Figure \ref{fig:dipi}(a) shows that there is a tight linear correlation between $b_{1,0}^{\rm Bi}$ and $b_{1,0}^{\rm Si}$, indicating that the fitted BMRs are well able to reproduce both the magnetic flux and the axial dipole moment of each individual SHARP. If the $a$  parameter in Equation (\ref{eqn:bmr}) is reduced from its optimum value of $0.56$, the linear relation remains but the slope reduces, as shown by the fainter points in Figure \ref{fig:dipi}(a).
Furthermore, as pointed out by \citet{wang1991}, $b_{1,0}^{\rm Bi}$ depends linearly on the flux, cosine-latitude, and latitudinal spread of the BMR \citep[cf.][]{jiang2014a}. Figure \ref{fig:dipi}(b) shows that a similar relationship holds for our BMRs.

Finally, summing $b_{1,0}^{\rm Si}$ over all regions gives a net axial dipole input of $3.32\,\mathrm{G}$, with total positive/negative contributions of $2.15\,\mathrm{G}$/$-0.64\,\mathrm{G}$ in the Northern hemisphere and $2.40\,\mathrm{G}$/$-0.59\,\mathrm{G}$ in the Southern hemisphere. Thus we estimate a slightly higher total dipole input than \citet{virtanen2019}, who found a net input of $2.91\,\mathrm{G}$ for Cycle 24 when they extracted active regions from a combination of NSO/SOLIS and HMI synoptic maps. We conjecture that the slightly higher total may relate to more flux being included in our extraction technique using SHARP data. On the other hand, we reproduce the trends found by \citet{virtanen2019} whereby, for Cycle 24, the Southern hemisphere has a stronger positive dipole input but a weaker negative dipole input than the Northern hemisphere. In Section \ref{s:predicted}, we consider how these initial dipole moments would evolve over the solar cycle as the active regions spread out and decay.

\section{Predicted asymptotic contributions} \label{s:predicted}

The contribution of an active region to the polar field at Solar Minimum is determined not by its axial dipole moment at emergence time, but by its eventual contribution to the axial dipole at Solar Minimum \citep[\textit{e.g.,}][]{wang1991,mackay2002}. In this section, we use surface flux transport (SFT)  modelling to predict this net contribution for each region in our database. In particular, we consider how the bipolar approximation affects the accuracy of this prediction.

\subsection{Surface flux transport model}\label{s:model}

We evolve $B(s,\phi,t)$ forward in time using the SFT model, which in our coordinates $s\equiv\sin\lambda$ and $\phi$ reads
\begin{equation}
\frac{\partial B}{\partial t} = \frac{D}{R_\odot^2}\left[\frac{\partial}{\partial s}\left((1-s^2)\frac{\partial B}{\partial s}\right) + \frac{1}{1-s^2}\frac{\partial^2B}{\partial\phi^2}\right] - \frac{\partial}{\partial s}\left[\frac{v_s(s)}{R_\odot}\sqrt{1-s^2} B\right]- \Omega(s)\frac{\partial B}{\partial\phi}.
\label{eqn:sft2d}
\end{equation}
Here $D$ is the supergranular diffusivity, $v_s(s)$ the meridional flow speed, and $\Omega(s)$ the angular velocity of differential rotation. In fact, it is not necessary to solve the two-dimensional equation, since $b_{1,0}$ depends only on the longitude-averaged field \citep{devore1984,cameron2007,iijima2017,petrovay2019}. Specifically, writing $\overline{B}(s,t)=(2\pi)^{-1}\int_0^{2\pi}B(s,\phi,t)\,\mathrm{d}\phi$, we have
\begin{equation}
b_{1,0}(t) = \frac{3}{2}\int_{-1}^1 s\overline{B}(s,t)\,\mathrm{d}s.
\end{equation}
Taking the longitude-average of Equation (\ref{eqn:sft2d}) leaves the evolution equation
\begin{equation}
\frac{\partial\overline{B}}{\partial t} = \frac{\partial}{\partial s}\left[\frac{D}{R_\odot^2}(1-s^2)\frac{\partial\overline{B}}{\partial s} - \frac{v_s(s)}{R_\odot}\sqrt{1-s^2}\overline{B}\right],
\label{eqn:evol}
\end{equation}
which we solve numerically using a finite-volume method on our computational grid. We use the meridional flow profile
\begin{equation}
v_s(s) = D_us(1-s^2)^{p/2},
\label{eqn:merid}
\end{equation}
as used in the optimisation exercise of \citet{whitbread2018}. We adopt their meridional flow profile shape of $p=2.33$, but not their values of $D=466.8\,\mathrm{km}^2\mathrm{s}^{-1}$ or $D_u = 0.025\,\mathrm{km}\,\mathrm{s}^{-1}$. This is because these values were fitted to Cycles 21-23, and we find them to produce too strong a polar field with our Cycle 24 input data. We found a better match of the observed evolution -- particularly of $b_{1,0}$ -- with  $D$ reduced to $350\,\mathrm{km}^2\mathrm{s}^{-1}$ and $D_u$ increased to $0.041\,\mathrm{km}\,\mathrm{s}^{-1}$. The latter value  corresponds to a peak poleward flow speed of $15\,\mathrm{m}\,\mathrm{s}^{-1}$ (see Appendix \ref{a:analytical}). We did not include an exponential decay time as this did not seem to be necessary to obtain a reasonable match to the observed evolution.

\subsection{Complete simulation}\label{s:complete}

\begin{figure}
\includegraphics[width=\textwidth]{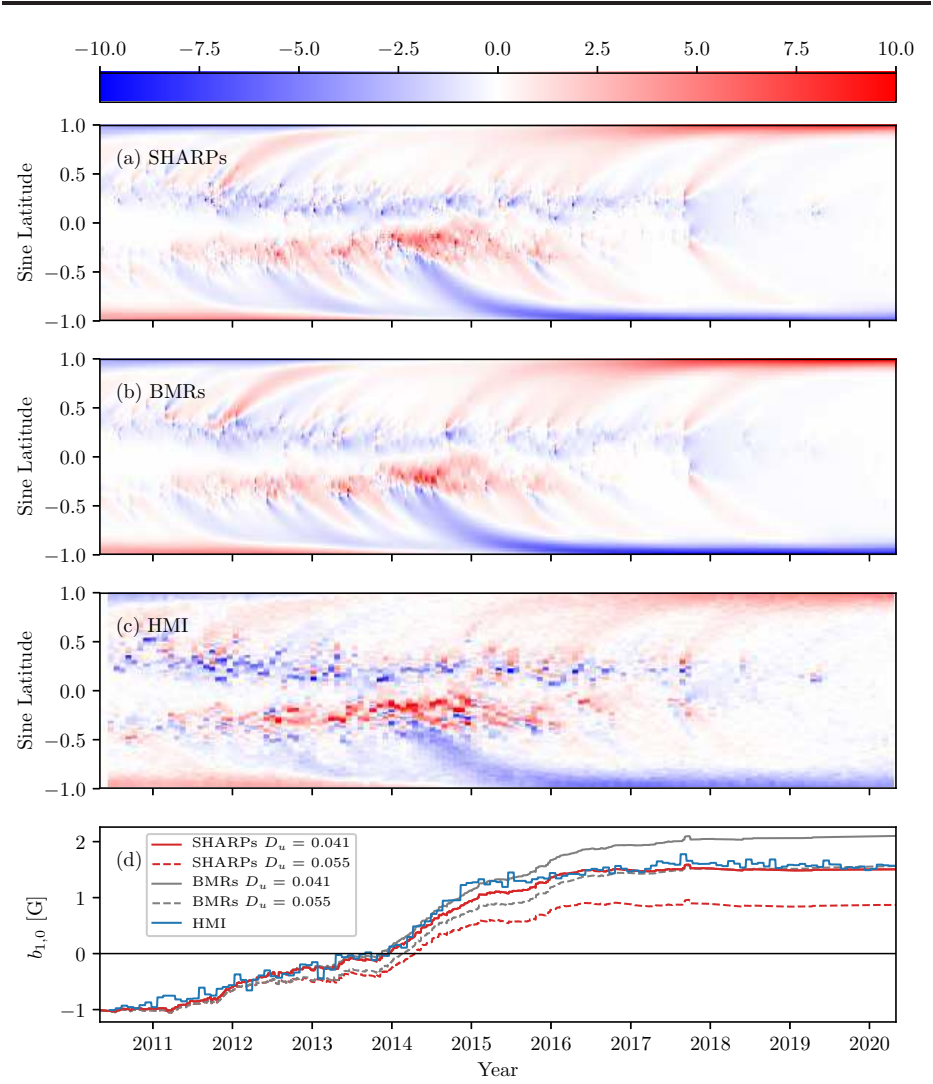}
\caption{Full SFT simulations. Panels (a) and (b) show $\overline{B}$ as a function of time and latitude, capped at $\pm 10\,{\rm G}$, for (a) the simulation with SHARP regions, and (b)  approximating BMRs. Panel (c) shows an observed supersynoptic map constructed from HMI data (see text), while panel (d) shows the axial dipole moments for the simulations and the HMI data (computed from panel c, with the early part being in good agreement with \citet{sun2015}). Dashed lines show simulations with increased $D_u$ (peak flow speed $20\,\mathrm{km}\,\mathrm{s}^{-1}$).}
\label{fig:all}
\end{figure}

First, we illustrate the SFT model with all regions included. Figures \ref{fig:all}(a) and (b) show the resulting time-latitude distributions of $\overline{B}$ in the SFT model where the sources are, respectively, the original SHARPs (mapped to the computational grid and with flux corrected, as in Figure \ref{fig:harp}c) or the fitted BMRs. For comparison, Figure \ref{fig:all}(c) shows an observed supersynoptic map determined from HMI data. Specifically, we use the radial component, pole-filled synoptic maps in the \texttt{hmi.synoptic\_mr\_polfil\_720s} series \citep{sun2018}. A smoothing filter of the form $\mathrm{e}^{-b_0l(l+1)}$ is applied to the spherical harmonic coefficients (with $b_0=10^{-4}$) before mapping to the our computational grid and correcting the flux balance multiplicatively. It is evident that both models do a reasonable job of matching the observed evolution of the large-scale field outside of active regions. Indeed, it is quite difficult to distinguish the two simulations visually (but compare, for example, the year 2014). 

If we plot the axial dipole moment, $b_{1,0}$, as in Figure \ref{fig:all}(d), then the difference between the two simulations becomes clearer. The solid lines show simulations with our standard parameters, chosen so that the SHARPs simulation best reproduces the observed evolution of $b_{1,0}$. Although $b_{1,0}$ from the BMR simulation remains close until the time of dipole reversal in 2014, it later increases, and overestimates the axial dipole in 2020 substantially. Specifically, the net change in $b_{1,0}$ over the whole simulation is $+2.51\,{\rm G}$ for the SHARPs but $+3.11\,{\rm G}$ for the BMRs. This amounts to a $24\%$ overestimate by the BMR simulation. Conversely, if the meridional flow speed is chosen so that the BMR simulation matches the observed axial dipole -- as shown by the dashed lines in Figure \ref{fig:all}(d) -- then the dipole reversal is delayed by several months and the evolution of $b_{1,0}$ is a poorer match for the observations between 2012 and 2017. To investigate why the BMR simulation overestimates the dipole moment, we next consider the evolution of the individual regions.

\subsection{Individual regions}\label{s:indiv}

\begin{figure}
\includegraphics[width=\textwidth]{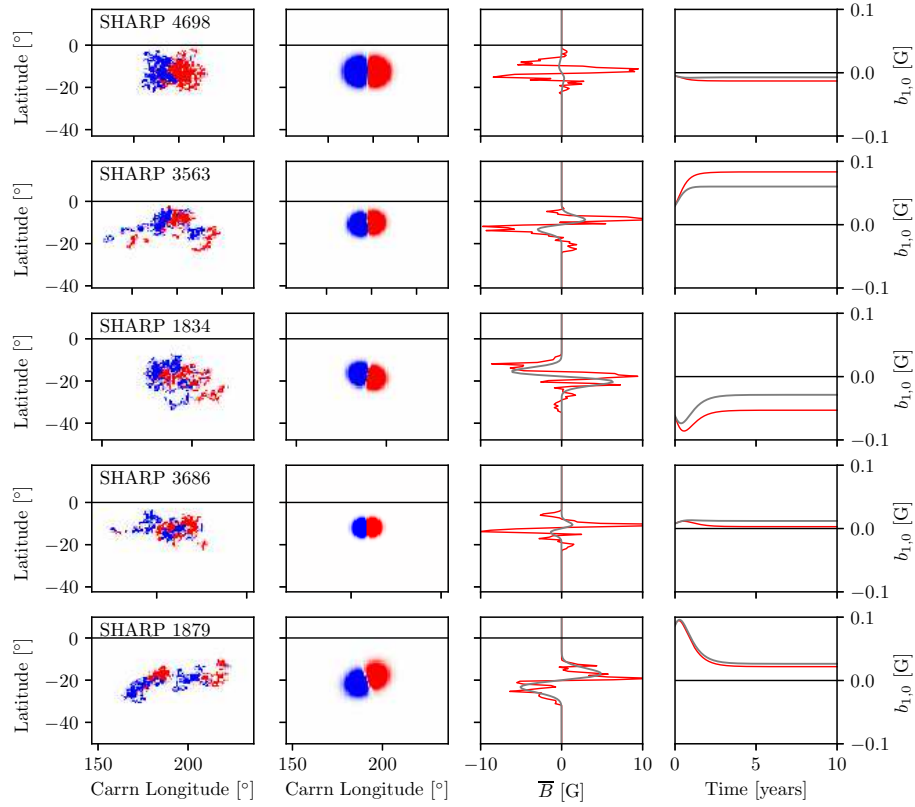}
\caption{The 5 regions with largest $|\Phi|$, showing their SHARPs (first column), approximating BMRs (second column), longitudinal averages (third column) and predicted evolution of their axial dipole moment with the SFT model (fourth column). In the third and fourth columns the red line shows the SHARP and the grey line the BMR.}
\label{fig:biggest}
\end{figure}

Since the SFT equation (\ref{eqn:evol}) is linear, we can determine the dipole contribution of an individual active region by simulating that region alone \citep{yeates2015}.
When it is initialised with a single BMR and there is no further emergence, the $\overline{B}$ distribution will evolve toward an asymptotic steady state satisfying
\begin{equation}
D(1-s^2)\frac{\partial\overline{B}}{\partial s} - v_s(s)\sqrt{1-s^2}\overline{B} = \mathrm{constant}.
\label{eqn:balance}
\end{equation}
There is an exact solution of this equation giving the latitudinal profile (see Appendix \ref{a:analytical}), but here we are interested in the magnitude of this asymptotic state, which we must compute numerically for each of the 1090 regions by evolving $\overline{B}$ forward in time for 10 years through Equation (\ref{eqn:evol}). This is ample time to reach the asymptotic profile, which typically takes only 2-3 years for these SFT parameters. From this asymptotic $\overline{B}$ profile, we compute the ``final'' axial dipole moment $b_{1,0}^{Bf}$ for each region numerically. As an example of this evolution, the rightmost column of Figure \ref{fig:biggest} shows the time evolution of $b_{1,0}$ for the five largest regions (by flux). For each region, we simulate both the evolution of the SHARP (first column) and of the fitted BMR (second column). For SHARPs 4698 and 1879, the dipole evolves in a similar way for both the SHARP and the approximating BMR. But for the other three regions in Figure \ref{fig:biggest}, there are significant discrepancies.

\begin{figure}
\includegraphics[width=0.5\textwidth]{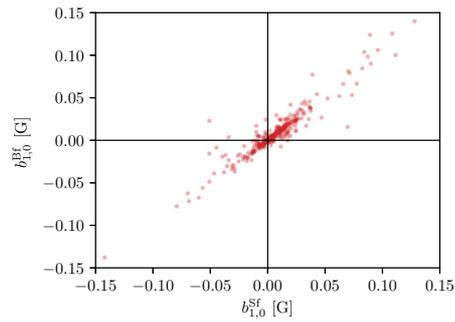}
\caption{Scatterplot of predicted axial dipole moments after 10-year evolution of the BMRs versus the corresponding SHARPs.}
\label{fig:dipf}
\end{figure}

Figure \ref{fig:dipf} shows a scatter plot of the predicted axial dipole moments using the BMRs versus the original SHARPs.  Overall there is a strong correlation, but with some scatter. The slight bias for $b_{1,0}^{\rm Bf} > b_{1,0}^{\rm Sf}$ that leads to the overestimate in the complete simulation is present but difficult to discern by eye from this figure.

\begin{figure}
\includegraphics[width=\textwidth]{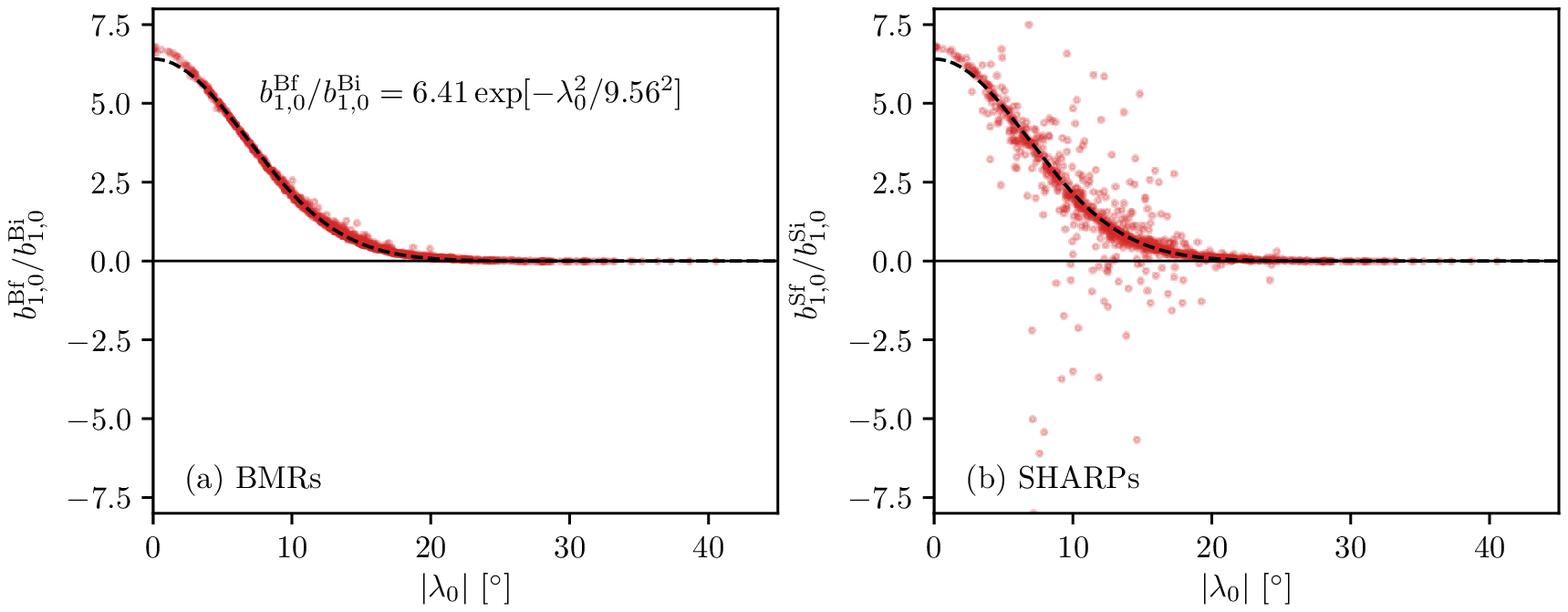}
\caption{Dipole amplification by the 10-year evolution for (a) the BMR approximation, and (b) the SHARPs, shown as a function of absolute latitude of the initial region. A least-squares fit to (a) is shown in both panels. In (b) there are a small number of extreme outliers not shown: the actual minimum value is $-343.8$ and the maximum is $75.3$.}
\label{fig:gaussian}
\end{figure}

In the SFT model, the dipole ``amplification factor'' for a BMR -- in other words, the ratio $b_{1,0}^{\rm Bf}/b_{1,0}^{\rm Bi}$ -- is known to be a Gaussian function of emergence  latitude. This was first noted by \citet{jiang2014a} and explained mathematically by \citet{petrovay2020a}. Figure \ref{fig:gaussian}(a) shows that our BMR-driven simulations do indeed follow this pattern. On the other hand, when we plot the same ratio for the SHARP regions, as in Figure \ref{fig:gaussian}(b), there is still a Gaussian pattern but now considerable scatter. Similar scatter was found by \citet{whitbread2018} who considered active regions from NSO data without making a BMR approximation.  For clarity, outliers are not shown in Figure \ref{fig:gaussian}, and when these are included the average ratio $b_{1,0}^{\rm Sf}/b_{1,0}^{\rm Si}$ at each latitude falls below the Gaussian fit from the BMR simulation. Again, this reflects the tendency of the BMR-driven simulation to overestimate the asymptotic dipole moment.

\begin{figure}
\includegraphics[width=\textwidth]{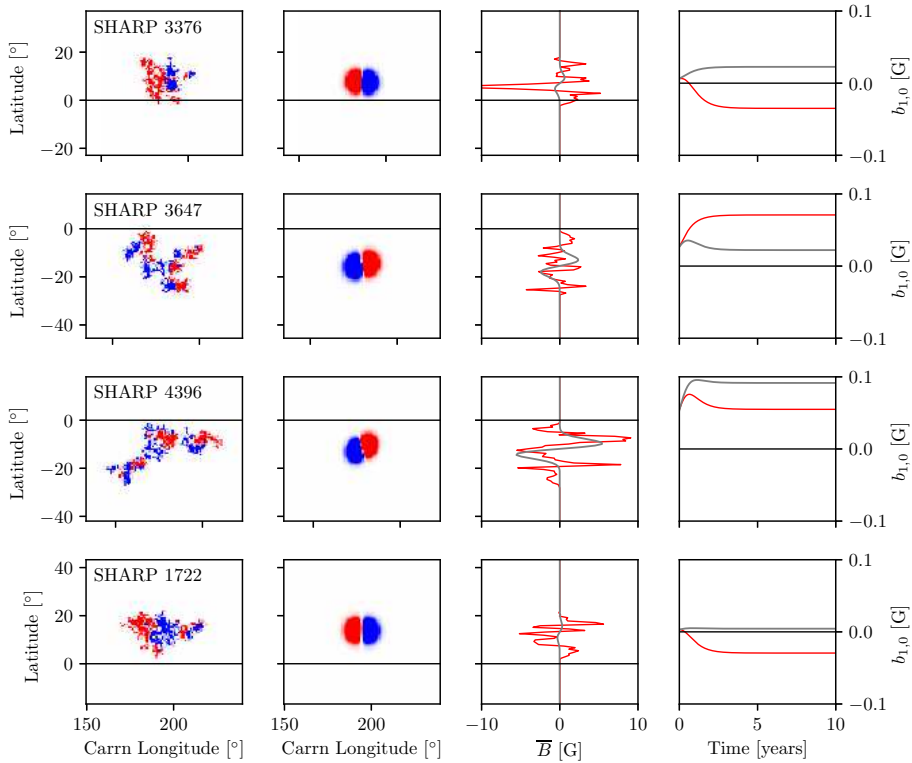}
\caption{The 4 regions with largest disparity $|b_{1,0}^{\rm Bf}-b_{1,0}^{\rm Sf}|$ in the predicted asymptotic dipole moment, showing their SHARPs (first column), approximating BMRs (second column), longitudinal averages (third column) and predicted evolution of their axial dipole moment (fourth column). In the third and fourth columns the red line shows the SHARP and the grey line the BMR.}
\label{fig:worst}
\end{figure}

To understand why BMRs tend, on average, to overestimate the asymptotic  dipole moment despite matching the initial dipole moment, it is instructive to look at the four regions with the greatest disparity, shown in Figure \ref{fig:worst}. From the first column, it is evident that the shapes of these regions diverge significantly from symmetric BMRs. 
In fact, it is only the longitude-average, $\overline{B}$, that matters, since this entirely determines the evolution of $b_{1,0}$. Accordingly, the third columns of Figures \ref{fig:biggest} and \ref{fig:worst} show the initial $\overline{B}$ profiles for both the SHARP (in red) and the BMR (in grey).
For example, SHARP 1879 in Figure \ref{fig:biggest} is well predicted by the BMR since the initial $\overline{B}$ is close to that of the SHARP, even though the original SHARP consists of (at least) two separate bipolar regions.
On the other hand, all four regions in Figure \ref{fig:worst} differ significantly from a symmetric BMR, even from the viewpoint of $\overline{B}$, leading to the disparity in predicted $b_{1,0}$ evolution. 

As discussed by \citet{iijima2019}, one way this can happen is if one polarity is more diffuse than the other; an example is SHARP 3376. Those authors argued that the effect will weaken the asymptotic dipole moment compared to the BMR, but here the effect is strong enough to reverse the sign. Overall, we find that the SHARP and BMR models predict final dipole moments of differing sign in 53 of our 1090 regions.
More generally, the evolution can differ if $\overline{B}$ for the SHARP contains significant variation that cannot be captured by a BMR, true for all of the regions in Figure \ref{fig:worst} (and several in Figure \ref{fig:biggest} too). \citet{jiang2019} presented a case study of just such a $\delta$-spot region where the predicted dipole is weaker than expected, again changing sign during the evolution because of the more complex initial shape. 
In other cases, the SHARP predicts essentially zero dipole moment while the BMR predicts a non-zero value, or \textit{vice versa} as in SHARP 1722.

\begin{figure}
\includegraphics[width=\textwidth]{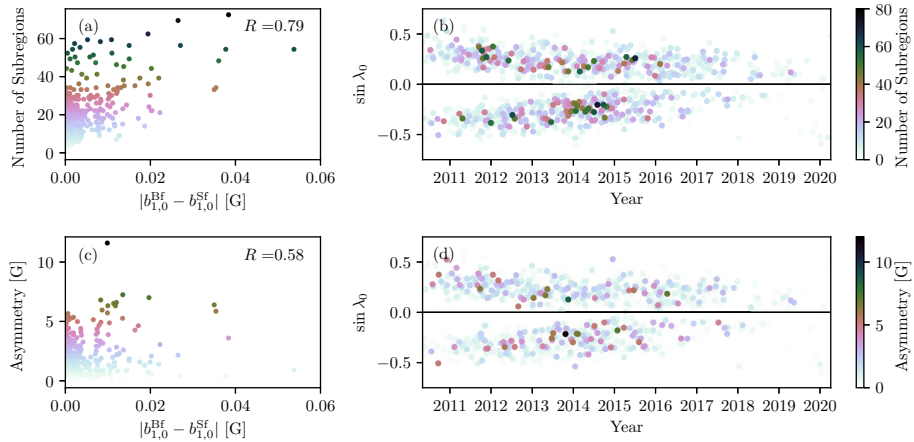}
\caption{Structure of the SHARPS, as measured by either the number of connected subregions with $|B^{i,j}|>100\,\mathrm{G}$ within each SHARP (top), or the asymmetry $\left|\max|\overline{B}^i| - \min|\overline{B}^i|\right|$ (bottom). The left column (a and c) shows each measure plotted against the disparity in predicted axial dipole, with $R$ showing the Spearman rank correlation coefficient (with negligible $p$ values). The right column (b and d) shows the SHARPs coloured by each measure as a function of latitude and time.}
\label{fig:disparity}
\end{figure}

Finally, Figure \ref{fig:disparity} represents an attempt to quantify the idea that regions with the greatest disparity between $b_{1,0}^{\rm Bf}$ and $b_{1,0}^{\rm Sf}$ tend to have either complex structure or asymmetry in the original SHARP. Specifically, Figure \ref{fig:disparity}(a) shows --- as a function of the disparity --- the number of connected subregions with (absolute) field strength above $100\,\mathrm{G}$ in the SHARP, which is a measure of complexity. Similarly, Figure \ref{fig:disparity}(c) shows the difference between positive and negative peaks in the SHARP's $\overline{B}$ profile, which is a measure of asymmetry. In both cases there are significant correlations, although stronger for the complexity measure than for the asymmetry. In addition, Figure \ref{fig:disparity}(b) shows significant clusters of complex regions during the Solar Maximum 2014-2015, which is the period when the axial dipole moments from the SHARP and BMR simulations are seen to diverge in Figure \ref{fig:all}(d).

\section{Conclusions}\label{s:conc}

In summary, using a new BMR database derived from HMI/SHARP data, we find that the SFT simulation driven by BMRs overestimates the axial dipole at the end of the solar cycle by 24\%. By contrast, using the SHARP maps themselves to drive the SFT model allow it to match the observed axial dipole evolution with the same SFT parameters. 

Since the initial BMRs and SHARPs had exactly the same dipole moment, our results suggest that it is likely not the omission of active region inflows that is responsible for the overestimation of the axial dipole by BMR-driven models. In fact, one could argue that the effect of such inflows is already indirectly included in our SFT simulations, through our optimisation of the meridional flow speed. The observed meridional flow is found to be inversely proportional to solar activity \citep[cf.][]{hathaway2014}, consistent with the effect of active region inflows and likely explaining why we needed a faster flow speed for this Cycle 24 simulation compared the \citet{whitbread2018} model.

We saw in Figure \ref{fig:all}(d) that changing the meridional flow speed can give the correct axial dipole at the end of the BMR-driven simulation, but at the expense of unrealistic intermediate evolution (the dashed grey curve in Figure \ref{fig:all}d). In particular, this delays the axial dipole reversal by a few months. In their recent parameter study for the SFT model, where the combined source term is represented by a pair of flux rings, \citet{petrovay2019} found that an exponential decay term was required in order to avoid such a delay in the dipole reversal. Here we find that such a term is not required in order to reverse the dipole at the correct time, provided the original SHARPs are used rather than BMRs.

Could the accuracy of the SFT model be improved within the context of BMR sources? To mimic past simulations \citep[\textit{e.g.},][]{yeates2014}, we also tried a run with all of the BMR tilts reduced from $\tilde{\gamma}$ to $g\tilde{\gamma}$. With $g=0.82$, the resulting simulation reproduced the correct final dipole moment, but again at the expense of the intermediate evolution -- in fact, the resulting curve for $b_{1,0}$ was close to the dashed gray curve in Figure \ref{fig:all}(d). It is possible that improved results could be obtained by selectively reducing the tilt angles according to the complexity of individual regions, but this remains for future investigation. In fact, it may be more effective to look at ways of fitting highly asymmetric or multipolar SHARPs (such as those shown in Figure \ref{fig:worst}) with asymmetric BMRs, or indeed with multiple BMRs. The manual databases of \citet{wang1989} and \citet{yeatesdata} were able to fit multiple BMRs to the same active region where appropriate, but an objective, automated procedure for this remains to be developed.

%%%%%%%%%%%%%%%%%%%%%%%%%%%%%%%%%%%%%%%%%%%%%%%%%%%%%%%%%%%%%%%%%%%%%%%%%%%
%% Appendix
%
\appendix
\section{Coordinate rotation} \label{a:rotation}

\begin{figure}
\includegraphics[width=\textwidth]{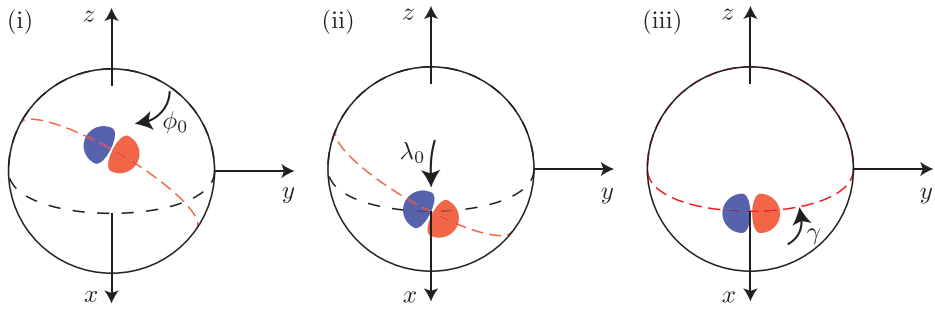}
\caption{The sequence of rotations to transform to coordinates where the BMR is at $s'=\phi'=0$ with $\gamma'=0$: (i) rotate angle $\phi_0$ around the $z$-axis; (ii) rotate angle $\lambda_0=\arcsin(s_0)$ around the (new) $y$-axis; and (iii) rotate angle $\gamma$ around the (new) $x$-axis.}
\label{fig:rot}
\end{figure}

Here we derive the coordinate transformation from the frame $(s,\phi)$ where the BMR is centred at $s=\phi=0$ to the frame $(s',\phi')$ where it is centred at $(s_0,\phi_0)$ with tilt $\gamma$. This amounts to a rotation, which is easiest to express in Cartesian coordinates
\begin{equation}
x = \cos\phi\sqrt{1-s^2}, \quad y=\sin\phi\sqrt{1-s^2},\quad z=s.
\end{equation}
Multiplying by the rotation matrices for the sequence of rotations indicated in Figure \ref{fig:rot} shows that Cartesian coordinates in the rotated frame are
\begin{eqnarray}
\left[
\begin{array}{c}
x'\\
y'\\
z'
\end{array}
\right] &=&
\left[
\begin{array}{ccc}
1\; & 0\; & 0\\
0\; & \cos\gamma\; & -\sin\gamma\\
0\; & \sin\gamma\; & \cos\gamma
\end{array}
\right]
\left[
\begin{array}{ccc}
\cos\lambda_0\; & 0\; & \sin\lambda_0\\
0\; & 1\; & 0\\
-\sin\lambda_0\; & 0\; & \cos\lambda_0
\end{array}
\right]
\cdot\\
&&\cdot
\left[
\begin{array}{ccc}
\cos\phi_0\; & \sin\phi_0\; & 0\\
-\sin\phi_0\; & \cos\phi_0\; & 0\\
0 & 0 & 1
\end{array}
\right]
\left[
\begin{array}{c}
x\\
y\\
z
\end{array}
\right],
\end{eqnarray}
where $s_0=\sin\lambda_0$. From these we determine $\phi'=\arctan(y'/x')$ and $s' = z'$.

\section{Analytical expressions} \label{a:analytical}

Here we note some analytical expressions for a meridional flow of the form (\ref{eqn:merid}), which readers may find useful \citep[see also][]{devore1984}. Firstly, one can show that the flow peaks at latitudes $s=\pm(1+p)^{-1/2}$ with speed
\begin{equation}
v_{\rm max} = \pm D_u p^{p/2}(1+p)^{-(1+p)/2}.
\end{equation}
Secondly, one can solve the ODE (\ref{eqn:balance}) exactly to find the shape of the asymptotic profile, which is
\begin{equation}
\overline{B}(s) = \overline{B}(1)\frac{\exp[-C(1-s^2)^{(1+p)/2}] - \mathrm{e}^{-C}}{1 - \mathrm{e}^{-C}}, \qquad C=\frac{D_uR_\odot}{(p+1)D}.
\label{eqn:asymb}
\end{equation}
Notice that the diffusivity $D$ enters only through the dimensionless parameter $C$. The actual amplitude of the profile, written here in terms of the value $\overline{B}(1)$ at the North pole, must be determined by following the evolution \citep[but for an approximate estimate of the asymptotic flux of a single active region, see][]{petrovay2020a}. Once the asymptotic value of $\overline{B}(1)$ is known, the axial dipole moment is found from (\ref{eqn:asymb}) to be
\begin{equation}
b_{1,0} = \frac{3\overline{B}_1}{2(1-\mathrm{e}^{-C})}\left[q\frac{ \Gamma(q) - \Gamma(q,C)}{C^q} - \mathrm{e}^{-C}\right],
\end{equation}
where $q=2(1+p)^{-1}$ and $\Gamma(q,C)$ is the incomplete Gamma function. In this paper, $b_{1,0}$ was simply computed numerically from the asymptotic solution $\overline{B}(s)$.

%%%%%%%%%%%%%%%%%%%%%%%%%%%%%%%%%%%%%%%%%%%%%%%%%%%%%%%%%%%%%%%%%%%%%%%%%%%
%% Acknowledgements
%
\begin{acks}
This work was supported by STFC grant ST/S000321/1.  The work benefited from discussions at the International Space Science Institute,  Bern, through the team on dynamo effectivity of solar active regions led by K. Petrovay. The author thanks P. Bhowmik and K. Petrovay for useful suggestions on an initial draft. The SDO data are courtesy of NASA and the SDO/HMI science team, and the NSO-derived BMR catalogues were downloaded from the solar dynamo dataverse (\url{https://dataverse.harvard.edu/dataverse/solardynamo}), maintained by A. Mu\~noz-Jaramillo.
\end{acks}

\section*{Code availability statement}

Our open-source Python code for BMR extraction is available online at \url{https://github.com/antyeates1983/sharps-bmrs}.

\section*{Data availability statement}

The database generated for this paper is available as a pre-generated ascii file on the Harvard Dataverse \citep{yeatesdata2020}. This may be reproduced with either the same or different parameters using the Python code.

%%% %%%%%%%%%%%%%%%%%%%%%%%%%%%%%%%%%%%%%%%%%%%%%%%%%%%%%%%%%%%
%% Bibliography
%
% Using BibTeX
%
\bibliographystyle{spr-mp-sola}
\bibliography{harps}  

\begin{thebibliography}{48}
% BibTex style file: spr-mp-sola.bst (nameyear), 2019-10-09
\ifx\bisbn     \undefined \def\bisbn  #1{ISBN #1}\fi
\ifx\binits    \undefined \def\binits#1{#1}\fi
\ifx\bauthor   \undefined \def\bauthor#1{#1}\fi
\ifx\batitle   \undefined \def\batitle#1{#1}\fi
\ifx\bjtitle   \undefined \def\bjtitle#1{\textit{#1}}\fi
\ifx\bvolume   \undefined \def\bvolume#1{\textbf{#1}}\fi
\ifx\byear     \undefined \def\byear#1{#1}\fi
\ifx\bissue    \undefined \def\bissue#1{#1}\fi
\ifx\bfpage    \undefined \def\bfpage#1{#1}\fi
\ifx\blpage    \undefined \def\blpage #1{#1}\fi
\ifx\burl      \undefined \def\burl#1{\textsf{#1}}\fi
\ifx\href      \undefined \def\href#1#2{\textsf{#2}}\fi
\ifx\betal     \undefined \def\betal{\textit{et al.}}\fi
\ifx\bctitle   \undefined \def\bctitle#1{#1}\fi
\ifx\beditor   \undefined \def\beditor#1{#1}\fi
\ifx\bbtitle   \undefined \def\bbtitle#1{\textit{#1}}\fi
\ifx\bedition  \undefined \def\bedition#1{#1}\fi
\ifx\bseriesno \undefined \def\bseriesno#1{\textbf{#1}}\fi
\ifx\blocation \undefined \def\blocation#1{#1}\fi
\ifx\bsertitle \undefined \def\bsertitle#1{\textit{#1}}\fi
\ifx\bsnm      \undefined \def\bsnm#1{#1}\fi
\ifx\bsuffix   \undefined \def\bsuffix#1{#1}\fi
\ifx\bparticle \undefined \def\bparticle#1{#1}\fi
\ifx\barticle  \undefined \def\barticle#1{}\fi
\ifx\binstitute  \undefined \def\binstitute#1{#1}\fi
\ifx\bpublisher  \undefined \def\bpublisher#1{#1}\fi
\ifx\doiurl    \undefined \def\doiurl#1{\href{#1}{\textsf{DOI}}}\fi
\makeatletter
\def\safeHref#1#2#3{\in@{http}{#2}\ifin@\href{#2}{#3}\else\href{#1#2}{#3}\fi}
\makeatother
\ifx\adsurl    \undefined
  \def\adsurl#1{\safeHref{https://ui.adsabs.harvard.edu/abs/}{#1}{\textsf{ADS}}}\fi
\ifx\arxivurl  \undefined
  \def\arxivurl#1{\safeHref{http://arxiv.org/abs/}{#1}{\textsf{arXiv}}}\fi
\ifx\botherref \undefined \def\botherref#1{}\fi
\ifx\url       \undefined \def\url#1{\textsf{#1}}\fi
\ifx\bchapter  \undefined \def\bchapter#1{}\fi
\ifx\bbook     \undefined \def\bbook#1{}\fi
\ifx\bcomment  \undefined \def\bcomment#1{#1}\fi
\ifx\oauthor   \undefined \def\oauthor#1{#1}\fi
\ifx\citeauthoryear \undefined\def \citeauthoryear#1{#1}\fi
\def\endbibitem {}
\ifx\bconflocation  \undefined \def\bconflocation#1{#1} \fi

\bibitem[\protect\citeauthoryear{{Bhowmik} and {Nandy}}{2018}]{bhowmik2018}
\begin{barticle}
\bauthor{\bsnm{{Bhowmik}}, \binits{P.}},
\bauthor{\bsnm{{Nandy}}, \binits{D.}}:
\byear{2018},
\batitle{{Prediction of the strength and timing of sunspot cycle 25 reveal
  decadal-scale space environmental conditions}}.
\bjtitle{Nature Comms.}
\bvolume{9},
\bfpage{5209}.
\doiurl{https://doi.org/10.1038/s41467-018-07690-0}.
\adsurl{2018NatCo...9.5209B}.
\end{barticle}
\endbibitem

\bibitem[\protect\citeauthoryear{{Bobra} \textit{et~al.}}{2014}]{Bobra2014}
\begin{barticle}
\bauthor{\bsnm{{Bobra}}, \binits{M.G.}},
\bauthor{\bsnm{{Sun}}, \binits{X.}},
\bauthor{\bsnm{{Hoeksema}}, \binits{J.T.}},
\bauthor{\bsnm{{Turmon}}, \binits{M.}},
\bauthor{\bsnm{{Liu}}, \binits{Y.}},
\bauthor{\bsnm{{Hayashi}}, \binits{K.}},
\bauthor{\bsnm{{Barnes}}, \binits{G.}},
\bauthor{\bsnm{{Leka}}, \binits{K.D.}}:
\byear{2014},
\batitle{{The Helioseismic and Magnetic Imager (HMI) Vector Magnetic Field
  Pipeline: SHARPs - Space-Weather HMI Active Region Patches}}.
\bjtitle{Solar Phys.}
\bvolume{289}(\bissue{9}),
\bfpage{3549}.
\doiurl{https://doi.org/10.1007/s11207-014-0529-3}.
\adsurl{2014SoPh..289.3549B}.
\end{barticle}
\endbibitem

\bibitem[\protect\citeauthoryear{{Cameron} and
  {Sch{\"u}ssler}}{2007}]{cameron2007}
\begin{barticle}
\bauthor{\bsnm{{Cameron}}, \binits{R.}},
\bauthor{\bsnm{{Sch{\"u}ssler}}, \binits{M.}}:
\byear{2007},
\batitle{{Solar Cycle Prediction Using Precursors and Flux Transport Models}}.
\bjtitle{Astrophys. J.}
\bvolume{659}(\bissue{1}),
\bfpage{801}.
\doiurl{https://doi.org/10.1086/512049}.
\adsurl{2007ApJ...659..801C}.
\end{barticle}
\endbibitem

\bibitem[\protect\citeauthoryear{{Cameron} \textit{et~al.}}{2010}]{cameron2010}
\begin{barticle}
\bauthor{\bsnm{{Cameron}}, \binits{R.H.}},
\bauthor{\bsnm{{Jiang}}, \binits{J.}},
\bauthor{\bsnm{{Schmitt}}, \binits{D.}},
\bauthor{\bsnm{{Sch{\"u}ssler}}, \binits{M.}}:
\byear{2010},
\batitle{{Surface Flux Transport Modeling for Solar Cycles 15-21: Effects of
  Cycle-Dependent Tilt Angles of Sunspot Groups}}.
\bjtitle{Astrophys. J.}
\bvolume{719}(\bissue{1}),
\bfpage{264}.
\doiurl{https://doi.org/10.1088/0004-637X/719/1/264}.
\adsurl{2010ApJ...719..264C}.
\end{barticle}
\endbibitem

\bibitem[\protect\citeauthoryear{{Cameron} \textit{et~al.}}{2013}]{cameron2013}
\begin{barticle}
\bauthor{\bsnm{{Cameron}}, \binits{R.H.}},
\bauthor{\bsnm{{Dasi-Espuig}}, \binits{M.}},
\bauthor{\bsnm{{Jiang}}, \binits{J.}},
\bauthor{\bsnm{{I{\textcommabelow s}{\i}k}}, \binits{E.}},
\bauthor{\bsnm{{Schmitt}}, \binits{D.}},
\bauthor{\bsnm{{Sch{\"u}ssler}}, \binits{M.}}:
\byear{2013},
\batitle{{Limits to solar cycle predictability: Cross-equatorial flux plumes}}.
\bjtitle{Astron. Astrophys.}
\bvolume{557},
\bfpage{A141}.
\doiurl{https://doi.org/10.1051/0004-6361/201321981}.
\adsurl{2013A&A...557A.141C}.
\end{barticle}
\endbibitem

\bibitem[\protect\citeauthoryear{{DeVore}, {Boris}, and
  {Sheeley}}{1984}]{devore1984}
\begin{barticle}
\bauthor{\bsnm{{DeVore}}, \binits{C.R.}},
\bauthor{\bsnm{{Boris}}, \binits{J.P.}},
\bauthor{\bsnm{{Sheeley}}, \binits{J.} \bsuffix{N.~R.}}:
\byear{1984},
\batitle{{The concentration of the large-scale solar magnetic field by a
  meridional surface flow}}.
\bjtitle{Solar Phys.}
\bvolume{92}(\bissue{1-2}),
\bfpage{1}.
\doiurl{https://doi.org/10.1007/BF00157230}.
\adsurl{1984SoPh...92....1D}.
\end{barticle}
\endbibitem

\bibitem[\protect\citeauthoryear{{Fisher} \textit{et~al.}}{2000}]{fisher2000}
\begin{barticle}
\bauthor{\bsnm{{Fisher}}, \binits{G.H.}},
\bauthor{\bsnm{{Fan}}, \binits{Y.}},
\bauthor{\bsnm{{Longcope}}, \binits{D.W.}},
\bauthor{\bsnm{{Linton}}, \binits{M.G.}},
\bauthor{\bsnm{{Pevtsov}}, \binits{A.A.}}:
\byear{2000},
\batitle{{The Solar Dynamo and Emerging Flux - (Invited Review)}}.
\bjtitle{Solar Phys.}
\bvolume{192},
\bfpage{119}.
\doiurl{https://doi.org/10.1023/A:1005286516009}.
\adsurl{2000SoPh..192..119F}.
\end{barticle}
\endbibitem

\bibitem[\protect\citeauthoryear{{Hathaway} and {Upton}}{2014}]{hathaway2014}
\begin{barticle}
\bauthor{\bsnm{{Hathaway}}, \binits{D.H.}},
\bauthor{\bsnm{{Upton}}, \binits{L.}}:
\byear{2014},
\batitle{{The solar meridional circulation and sunspot cycle variability}}.
\bjtitle{J. Geophys. Res. (Space Physi)}
\bvolume{119}(\bissue{5}),
\bfpage{3316}.
\doiurl{https://doi.org/10.1002/2013JA019432}.
\adsurl{2014JGRA..119.3316H}.
\end{barticle}
\endbibitem

\bibitem[\protect\citeauthoryear{{Hickmann}
  \textit{et~al.}}{2015}]{hickman2015}
\begin{barticle}
\bauthor{\bsnm{{Hickmann}}, \binits{K.S.}},
\bauthor{\bsnm{{Godinez}}, \binits{H.C.}},
\bauthor{\bsnm{{Henney}}, \binits{C.J.}},
\bauthor{\bsnm{{Arge}}, \binits{C.N.}}:
\byear{2015},
\batitle{{Data Assimilation in the ADAPT Photospheric Flux Transport Model}}.
\bjtitle{Solar Phys.}
\bvolume{290}(\bissue{4}),
\bfpage{1105}.
\doiurl{https://doi.org/10.1007/s11207-015-0666-3}.
\adsurl{2015SoPh..290.1105H}.
\end{barticle}
\endbibitem

\bibitem[\protect\citeauthoryear{{Iijima}, {Hotta}, and
  {Imada}}{2019}]{iijima2019}
\begin{barticle}
\bauthor{\bsnm{{Iijima}}, \binits{H.}},
\bauthor{\bsnm{{Hotta}}, \binits{H.}},
\bauthor{\bsnm{{Imada}}, \binits{S.}}:
\byear{2019},
\batitle{{Effect of Morphological Asymmetry between Leading and Following
  Sunspots on the Prediction of Solar Cycle Activity}}.
\bjtitle{Astrophys. J.}
\bvolume{883}(\bissue{1}),
\bfpage{24}.
\doiurl{https://doi.org/10.3847/1538-4357/ab3b04}.
\adsurl{2019ApJ...883...24I}.
\end{barticle}
\endbibitem

\bibitem[\protect\citeauthoryear{{Iijima} \textit{et~al.}}{2017}]{iijima2017}
\begin{barticle}
\bauthor{\bsnm{{Iijima}}, \binits{H.}},
\bauthor{\bsnm{{Hotta}}, \binits{H.}},
\bauthor{\bsnm{{Imada}}, \binits{S.}},
\bauthor{\bsnm{{Kusano}}, \binits{K.}},
\bauthor{\bsnm{{Shiota}}, \binits{D.}}:
\byear{2017},
\batitle{{Improvement of solar-cycle prediction: Plateau of solar axial dipole
  moment}}.
\bjtitle{Astron. Astrophys.}
\bvolume{607},
\bfpage{L2}.
\doiurl{https://doi.org/10.1051/0004-6361/201731813}.
\adsurl{2017A&A...607L...2I}.
\end{barticle}
\endbibitem

\bibitem[\protect\citeauthoryear{{Jiang}, {Cameron}, and
  {Sch{\"u}ssler}}{2014}]{jiang2014a}
\begin{barticle}
\bauthor{\bsnm{{Jiang}}, \binits{J.}},
\bauthor{\bsnm{{Cameron}}, \binits{R.H.}},
\bauthor{\bsnm{{Sch{\"u}ssler}}, \binits{M.}}:
\byear{2014},
\batitle{{Effects of the Scatter in Sunspot Group Tilt Angles on the
  Large-scale Magnetic Field at the Solar Surface}}.
\bjtitle{Astrophys. J.}
\bvolume{791}(\bissue{1}),
\bfpage{5}.
\doiurl{https://doi.org/10.1088/0004-637X/791/1/5}.
\adsurl{2014ApJ...791....5J}.
\end{barticle}
\endbibitem

\bibitem[\protect\citeauthoryear{{Jiang} \textit{et~al.}}{2014}]{jiang2014}
\begin{barticle}
\bauthor{\bsnm{{Jiang}}, \binits{J.}},
\bauthor{\bsnm{{Hathaway}}, \binits{D.H.}},
\bauthor{\bsnm{{Cameron}}, \binits{R.H.}},
\bauthor{\bsnm{{Solanki}}, \binits{S.K.}},
\bauthor{\bsnm{{Gizon}}, \binits{L.}},
\bauthor{\bsnm{{Upton}}, \binits{L.}}:
\byear{2014},
\batitle{{Magnetic Flux Transport at the Solar Surface}}.
\bjtitle{Space Sci. Rev.}
\bvolume{186}(\bissue{1-4}),
\bfpage{491}.
\doiurl{https://doi.org/10.1007/s11214-014-0083-1}.
\adsurl{2014SSRv..186..491J}.
\end{barticle}
\endbibitem

\bibitem[\protect\citeauthoryear{{Jiang} \textit{et~al.}}{2018}]{jiang2018}
\begin{barticle}
\bauthor{\bsnm{{Jiang}}, \binits{J.}},
\bauthor{\bsnm{{Wang}}, \binits{J.-X.}},
\bauthor{\bsnm{{Jiao}}, \binits{Q.-R.}},
\bauthor{\bsnm{{Cao}}, \binits{J.-B.}}:
\byear{2018},
\batitle{{Predictability of the Solar Cycle Over One Cycle}}.
\bjtitle{Astrophys. J.}
\bvolume{863}(\bissue{2}),
\bfpage{159}.
\doiurl{https://doi.org/10.3847/1538-4357/aad197}.
\adsurl{2018ApJ...863..159J}.
\end{barticle}
\endbibitem

\bibitem[\protect\citeauthoryear{{Jiang} \textit{et~al.}}{2019}]{jiang2019}
\begin{barticle}
\bauthor{\bsnm{{Jiang}}, \binits{J.}},
\bauthor{\bsnm{{Song}}, \binits{Q.}},
\bauthor{\bsnm{{Wang}}, \binits{J.-X.}},
\bauthor{\bsnm{{Baranyi}}, \binits{T.}}:
\byear{2019},
\batitle{{Different Contributions to Space Weather and Space Climate from
  Different Big Solar Active Regions}}.
\bjtitle{Astrophys. J.}
\bvolume{871}(\bissue{1}),
\bfpage{16}.
\doiurl{https://doi.org/10.3847/1538-4357/aaf64a}.
\adsurl{2019ApJ...871...16J}.
\end{barticle}
\endbibitem

\bibitem[\protect\citeauthoryear{{Labonville}, {Charbonneau}, and
  {Lemerle}}{2019}]{labonville2019}
\begin{barticle}
\bauthor{\bsnm{{Labonville}}, \binits{F.}},
\bauthor{\bsnm{{Charbonneau}}, \binits{P.}},
\bauthor{\bsnm{{Lemerle}}, \binits{A.}}:
\byear{2019},
\batitle{{A Dynamo-based Forecast of Solar Cycle 25}}.
\bjtitle{Solar Phys.}
\bvolume{294}(\bissue{6}),
\bfpage{82}.
\doiurl{https://doi.org/10.1007/s11207-019-1480-0}.
\adsurl{2019SoPh..294...82L}.
\end{barticle}
\endbibitem

\bibitem[\protect\citeauthoryear{{Leighton}}{1964}]{leighton1964}
\begin{barticle}
\bauthor{\bsnm{{Leighton}}, \binits{R.B.}}:
\byear{1964},
\batitle{{Transport of Magnetic Fields on the Sun.}}
\bjtitle{Astrophys. J.}
\bvolume{140},
\bfpage{1547}.
\doiurl{https://doi.org/10.1086/148058}.
\adsurl{1964ApJ...140.1547L}.
\end{barticle}
\endbibitem

\bibitem[\protect\citeauthoryear{{Mackay} and {Yeates}}{2012}]{mackay2012}
\begin{barticle}
\bauthor{\bsnm{{Mackay}}, \binits{D.H.}},
\bauthor{\bsnm{{Yeates}}, \binits{A.R.}}:
\byear{2012},
\batitle{{The Sun's Global Photospheric and Coronal Magnetic Fields:
  Observations and Models}}.
\bjtitle{Living Rev. Solar Phys.}
\bvolume{9}(\bissue{1}),
\bfpage{6}.
\doiurl{https://doi.org/10.12942/lrsp-2012-6}.
\adsurl{2012LRSP....9....6M}.
\end{barticle}
\endbibitem

\bibitem[\protect\citeauthoryear{{Mackay}, {Priest}, and
  {Lockwood}}{2002}]{mackay2002}
\begin{barticle}
\bauthor{\bsnm{{Mackay}}, \binits{D.H.}},
\bauthor{\bsnm{{Priest}}, \binits{E.R.}},
\bauthor{\bsnm{{Lockwood}}, \binits{M.}}:
\byear{2002},
\batitle{{The Evolution of the Sun's Open Magnetic Flux - I. A Single Bipole}}.
\bjtitle{Solar Phys.}
\bvolume{207}(\bissue{2}),
\bfpage{291}.
\doiurl{https://doi.org/10.1023/A:1016249917230}.
\adsurl{2002SoPh..207..291M}.
\end{barticle}
\endbibitem

\bibitem[\protect\citeauthoryear{{McClintock} and
  {Norton}}{2013}]{mcclintock2013}
\begin{barticle}
\bauthor{\bsnm{{McClintock}}, \binits{B.H.}},
\bauthor{\bsnm{{Norton}}, \binits{A.A.}}:
\byear{2013},
\batitle{{Recovering Joy's Law as a Function of Solar Cycle, Hemisphere, and
  Longitude}}.
\bjtitle{Solar Phys.}
\bvolume{287}(\bissue{1-2}),
\bfpage{215}.
\doiurl{https://doi.org/10.1007/s11207-013-0338-0}.
\adsurl{2013SoPh..287..215M}.
\end{barticle}
\endbibitem

\bibitem[\protect\citeauthoryear{{Mu{\~n}oz-Jaramillo}
  \textit{et~al.}}{2013}]{munoz2013}
\begin{barticle}
\bauthor{\bsnm{{Mu{\~n}oz-Jaramillo}}, \binits{A.}},
\bauthor{\bsnm{{Dasi-Espuig}}, \binits{M.}},
\bauthor{\bsnm{{Balmaceda}}, \binits{L.A.}},
\bauthor{\bsnm{{DeLuca}}, \binits{E.E.}}:
\byear{2013},
\batitle{{Solar Cycle Propagation, Memory, and Prediction: Insights from a
  Century of Magnetic Proxies}}.
\bjtitle{Astrophys. J. Lett.}
\bvolume{767}(\bissue{2}),
\bfpage{L25}.
\doiurl{https://doi.org/10.1088/2041-8205/767/2/L25}.
\adsurl{2013ApJ...767L..25M}.
\end{barticle}
\endbibitem

\bibitem[\protect\citeauthoryear{{Nagy} \textit{et~al.}}{2017}]{nagy2017}
\begin{barticle}
\bauthor{\bsnm{{Nagy}}, \binits{M.}},
\bauthor{\bsnm{{Lemerle}}, \binits{A.}},
\bauthor{\bsnm{{Labonville}}, \binits{F.}},
\bauthor{\bsnm{{Petrovay}}, \binits{K.}},
\bauthor{\bsnm{{Charbonneau}}, \binits{P.}}:
\byear{2017},
\batitle{{The Effect of ``Rogue'' Active Regions on the Solar Cycle}}.
\bjtitle{Solar Phys.}
\bvolume{292}(\bissue{11}),
\bfpage{167}.
\doiurl{https://doi.org/10.1007/s11207-017-1194-0}.
\adsurl{2017SoPh..292..167N}.
\end{barticle}
\endbibitem

\bibitem[\protect\citeauthoryear{{Pesnell}}{2016}]{pesnell2016}
\begin{barticle}
\bauthor{\bsnm{{Pesnell}}, \binits{W.D.}}:
\byear{2016},
\batitle{{Predictions of Solar Cycle 24: How are we doing?}}
\bjtitle{Space Weather}
\bvolume{14}(\bissue{1}),
\bfpage{10}.
\doiurl{https://doi.org/10.1002/2015SW001304}.
\adsurl{2016SpWea..14...10P}.
\end{barticle}
\endbibitem

\bibitem[\protect\citeauthoryear{{Petrovay}}{2020}]{petrovay2020}
\begin{barticle}
\bauthor{\bsnm{{Petrovay}}, \binits{K.}}:
\byear{2020},
\batitle{{Solar cycle prediction}}.
\bjtitle{Living Rev. Solar Phys.}
\bvolume{17}(\bissue{1}),
\bfpage{2}.
\doiurl{https://doi.org/10.1007/s41116-020-0022-z}.
\adsurl{2020LRSP...17....2P}.
\end{barticle}
\endbibitem

\bibitem[\protect\citeauthoryear{{Petrovay} and {Talafha}}{2019}]{petrovay2019}
\begin{barticle}
\bauthor{\bsnm{{Petrovay}}, \binits{K.}},
\bauthor{\bsnm{{Talafha}}, \binits{M.}}:
\byear{2019},
\batitle{{Optimization of surface flux transport models for the solar polar
  magnetic field}}.
\bjtitle{Astron. Astrophys.}
\bvolume{632},
\bfpage{A87}.
\doiurl{https://doi.org/10.1051/0004-6361/201936099}.
\adsurl{2019A&A...632A..87P}.
\end{barticle}
\endbibitem

\bibitem[\protect\citeauthoryear{{Petrovay}, {Nagy}, and
  {Yeates}}{2020}]{petrovay2020a}
\begin{botherref}
\oauthor{\bsnm{{Petrovay}}, \binits{K.}},
\oauthor{\bsnm{{Nagy}}, \binits{M.}},
\oauthor{\bsnm{{Yeates}}, \binits{A.R.}}:
2020,
{Towards an algebraic method of solar cycle prediction. I. Calculating the
  ultimate dipole contributions of individual active regions},
in preparation.
\end{botherref}
\endbibitem

\bibitem[\protect\citeauthoryear{{Schatten}
  \textit{et~al.}}{1978}]{schatten1978}
\begin{barticle}
\bauthor{\bsnm{{Schatten}}, \binits{K.H.}},
\bauthor{\bsnm{{Scherrer}}, \binits{P.H.}},
\bauthor{\bsnm{{Svalgaard}}, \binits{L.}},
\bauthor{\bsnm{{Wilcox}}, \binits{J.M.}}:
\byear{1978},
\batitle{{Using Dynamo Theory to predict the sunspot number during Solar Cycle
  21}}.
\bjtitle{Geophys. Res. Lett.}
\bvolume{5}(\bissue{5}),
\bfpage{411}.
\doiurl{https://doi.org/10.1029/GL005i005p00411}.
\adsurl{1978GeoRL...5..411S}.
\end{barticle}
\endbibitem

\bibitem[\protect\citeauthoryear{{Sheeley}}{2005}]{sheeley2005}
\begin{barticle}
\bauthor{\bsnm{{Sheeley}}, \binits{N.R.}}:
\byear{2005},
\batitle{{Surface Evolution of the Sun's Magnetic Field: A Historical Review of
  the Flux-Transport Mechanism}}.
\bjtitle{Living Rev.Solar Phys.}
\bvolume{2}(\bissue{1}),
\bfpage{5}.
\doiurl{https://doi.org/10.12942/lrsp-2005-5}.
\adsurl{2005LRSP....2....5S}.
\end{barticle}
\endbibitem

\bibitem[\protect\citeauthoryear{{Sheeley} and {Wang}}{2016}]{wangdata}
\begin{botherref}
\oauthor{\bsnm{{Sheeley}}, \binits{N.R.}},
\oauthor{\bsnm{{Wang}}, \binits{Y.-M.}}:
2016,
{Bipolar Magnetic Regions determined from Kitt Peak Vacuum Telescope
  Magnetograms}.
\doiurl{https://doi.org/10.7910/DVN/TF6TY4}.
\url{https://doi.org/10.7910/DVN/TF6TY4}.
\end{botherref}
\endbibitem

\bibitem[\protect\citeauthoryear{{Stenflo} and
  {Kosovichev}}{2012}]{stenflo2012}
\begin{barticle}
\bauthor{\bsnm{{Stenflo}}, \binits{J.O.}},
\bauthor{\bsnm{{Kosovichev}}, \binits{A.G.}}:
\byear{2012},
\batitle{{Bipolar Magnetic Regions on the Sun: Global Analysis of the SOHO/MDI
  Data Set}}.
\bjtitle{Astrophys. J.}
\bvolume{745}(\bissue{2}),
\bfpage{129}.
\doiurl{https://doi.org/10.1088/0004-637X/745/2/129}.
\adsurl{2012ApJ...745..129S}.
\end{barticle}
\endbibitem

\bibitem[\protect\citeauthoryear{{Sun}}{2018}]{sun2018}
\begin{botherref}
\oauthor{\bsnm{{Sun}}, \binits{X.}}:
2018,
{Polar Field Correction for HMI Line-of-Sight Synoptic Data}.
\textit{arXiv e-prints},
arXiv:1801.04265.
\adsurl{2018arXiv180104265S}.
\end{botherref}
\endbibitem

\bibitem[\protect\citeauthoryear{{Sun} \textit{et~al.}}{2015}]{sun2015}
\begin{barticle}
\bauthor{\bsnm{{Sun}}, \binits{X.}},
\bauthor{\bsnm{{Hoeksema}}, \binits{J.T.}},
\bauthor{\bsnm{{Liu}}, \binits{Y.}},
\bauthor{\bsnm{{Zhao}}, \binits{J.}}:
\byear{2015},
\batitle{{On Polar Magnetic Field Reversal and Surface Flux Transport During
  Solar Cycle 24}}.
\bjtitle{Astrophys. J.}
\bvolume{798}(\bissue{2}),
\bfpage{114}.
\doiurl{https://doi.org/10.1088/0004-637X/798/2/114}.
\adsurl{2015ApJ...798..114S}.
\end{barticle}
\endbibitem

\bibitem[\protect\citeauthoryear{{Tlatova} \textit{et~al.}}{2018}]{tlatova2018}
\begin{barticle}
\bauthor{\bsnm{{Tlatova}}, \binits{K.}},
\bauthor{\bsnm{{Tlatov}}, \binits{A.}},
\bauthor{\bsnm{{Pevtsov}}, \binits{A.}},
\bauthor{\bsnm{{Mursula}}, \binits{K.}},
\bauthor{\bsnm{{Vasil'eva}}, \binits{V.}},
\bauthor{\bsnm{{Heikkinen}}, \binits{E.}},
\bauthor{\bsnm{{Bertello}}, \binits{L.}},
\bauthor{\bsnm{{Pevtsov}}, \binits{A.}},
\bauthor{\bsnm{{Virtanen}}, \binits{I.}},
\bauthor{\bsnm{{Karachik}}, \binits{N.}}:
\byear{2018},
\batitle{{Tilt of Sunspot Bipoles in Solar Cycles 15 to 24}}.
\bjtitle{Solar Phys.}
\bvolume{293}(\bissue{8}),
\bfpage{118}.
\doiurl{https://doi.org/10.1007/s11207-018-1337-y}.
\adsurl{2018SoPh..293..118T}.
\end{barticle}
\endbibitem

\bibitem[\protect\citeauthoryear{{Upton} and {Hathaway}}{2014}]{upton2014}
\begin{barticle}
\bauthor{\bsnm{{Upton}}, \binits{L.}},
\bauthor{\bsnm{{Hathaway}}, \binits{D.H.}}:
\byear{2014},
\batitle{{Predicting the Sun's Polar Magnetic Fields with a Surface Flux
  Transport Model}}.
\bjtitle{Astrophys. J.}
\bvolume{780}(\bissue{1}),
\bfpage{5}.
\doiurl{https://doi.org/10.1088/0004-637X/780/1/5}.
\adsurl{2014ApJ...780....5U}.
\end{barticle}
\endbibitem

\bibitem[\protect\citeauthoryear{{Upton} and {Hathaway}}{2018}]{upton2018}
\begin{barticle}
\bauthor{\bsnm{{Upton}}, \binits{L.A.}},
\bauthor{\bsnm{{Hathaway}}, \binits{D.H.}}:
\byear{2018},
\batitle{{An Updated Solar Cycle 25 Prediction With AFT: The Modern Minimum}}.
\bjtitle{Geophys. Res. Lett.}
\bvolume{45}(\bissue{16}),
\bfpage{8091}.
\doiurl{https://doi.org/10.1029/2018GL078387}.
\adsurl{2018GeoRL..45.8091U}.
\end{barticle}
\endbibitem

\bibitem[\protect\citeauthoryear{{van Driel-Gesztelyi} and
  {Green}}{2015}]{vandriel2015}
\begin{barticle}
\bauthor{\bsnm{{van Driel-Gesztelyi}}, \binits{L.}},
\bauthor{\bsnm{{Green}}, \binits{L.M.}}:
\byear{2015},
\batitle{{Evolution of Active Regions}}.
\bjtitle{Living Rev. Solar Phys.}
\bvolume{12}(\bissue{1}),
\bfpage{1}.
\doiurl{https://doi.org/10.1007/lrsp-2015-1}.
\adsurl{2015LRSP...12....1V}.
\end{barticle}
\endbibitem

\bibitem[\protect\citeauthoryear{{Virtanen}
  \textit{et~al.}}{2019a}]{virtanen2019p4}
\begin{barticle}
\bauthor{\bsnm{{Virtanen}}, \binits{I.O.I.}},
\bauthor{\bsnm{{Virtanen}}, \binits{I.I.}},
\bauthor{\bsnm{{Pevtsov}}, \binits{A.A.}},
\bauthor{\bsnm{{Bertello}}, \binits{L.}},
\bauthor{\bsnm{{Yeates}}, \binits{A.}},
\bauthor{\bsnm{{Mursula}}, \binits{K.}}:
\byear{2019}a,
\batitle{{Reconstructing solar magnetic fields from historical observations.
  IV. Testing the reconstruction method}}.
\bjtitle{Astron. Astrophys.}
\bvolume{627},
\bfpage{A11}.
\doiurl{https://doi.org/10.1051/0004-6361/201935606}.
\adsurl{2019A&A...627A..11V}.
\end{barticle}
\endbibitem

\bibitem[\protect\citeauthoryear{{Virtanen}
  \textit{et~al.}}{2019b}]{virtanen2019}
\begin{barticle}
\bauthor{\bsnm{{Virtanen}}, \binits{I.O.I.}},
\bauthor{\bsnm{{Virtanen}}, \binits{I.I.}},
\bauthor{\bsnm{{Pevtsov}}, \binits{A.A.}},
\bauthor{\bsnm{{Mursula}}, \binits{K.}}:
\byear{2019}b,
\batitle{{Reconstructing solar magnetic fields from historical observations.
  VI. Axial dipole moments of solar active regions in cycles 21-24}}.
\bjtitle{Astron. Astrophys.}
\bvolume{632},
\bfpage{A39}.
\doiurl{https://doi.org/10.1051/0004-6361/201936134}.
\adsurl{2019A&A...632A..39V}.
\end{barticle}
\endbibitem

\bibitem[\protect\citeauthoryear{{Wang}}{2017}]{wang2017}
\begin{barticle}
\bauthor{\bsnm{{Wang}}, \binits{Y.-M.}}:
\byear{2017},
\batitle{{Surface Flux Transport and the Evolution of the Sun's Polar Fields}}.
\bjtitle{Space Sci. Rev.}
\bvolume{210}(\bissue{1-4}),
\bfpage{351}.
\doiurl{https://doi.org/10.1007/s11214-016-0257-0}.
\adsurl{2017SSRv..210..351W}.
\end{barticle}
\endbibitem

\bibitem[\protect\citeauthoryear{{Wang} and {Sheeley}}{1989}]{wang1989}
\begin{barticle}
\bauthor{\bsnm{{Wang}}, \binits{Y.-M.}},
\bauthor{\bsnm{{Sheeley}}, \binits{J.} \bsuffix{N.~R.}}:
\byear{1989},
\batitle{{Average Properties of Bipolar Magnetic Regions during Sunspot
  CYCLE-21}}.
\bjtitle{Solar Phys.}
\bvolume{124}(\bissue{1}),
\bfpage{81}.
\doiurl{https://doi.org/10.1007/BF00146521}.
\adsurl{1989SoPh..124...81W}.
\end{barticle}
\endbibitem

\bibitem[\protect\citeauthoryear{{Wang} and {Sheeley}}{1991}]{wang1991}
\begin{barticle}
\bauthor{\bsnm{{Wang}}, \binits{Y.-M.}},
\bauthor{\bsnm{{Sheeley}}, \binits{J.} \bsuffix{N.~R.}}:
\byear{1991},
\batitle{{Magnetic Flux Transport and the Sun's Dipole Moment: New Twists to
  the Babcock-Leighton Model}}.
\bjtitle{Astrophys. J.}
\bvolume{375},
\bfpage{761}.
\doiurl{https://doi.org/10.1086/170240}.
\adsurl{1991ApJ...375..761W}.
\end{barticle}
\endbibitem

\bibitem[\protect\citeauthoryear{{Whitbread}, {Yeates}, and
  {Mu{\~n}oz-Jaramillo}}{2018}]{whitbread2018}
\begin{barticle}
\bauthor{\bsnm{{Whitbread}}, \binits{T.}},
\bauthor{\bsnm{{Yeates}}, \binits{A.R.}},
\bauthor{\bsnm{{Mu{\~n}oz-Jaramillo}}, \binits{A.}}:
\byear{2018},
\batitle{{How Many Active Regions Are Necessary to Predict the Solar Dipole
  Moment?}}
\bjtitle{Astrophys. J.}
\bvolume{863}(\bissue{2}),
\bfpage{116}.
\doiurl{https://doi.org/10.3847/1538-4357/aad17e}.
\adsurl{2018ApJ...863..116W}.
\end{barticle}
\endbibitem

\bibitem[\protect\citeauthoryear{{Whitbread}
  \textit{et~al.}}{2017}]{whitbread2017}
\begin{barticle}
\bauthor{\bsnm{{Whitbread}}, \binits{T.}},
\bauthor{\bsnm{{Yeates}}, \binits{A.R.}},
\bauthor{\bsnm{{Mu{\~n}oz-Jaramillo}}, \binits{A.}},
\bauthor{\bsnm{{Petrie}}, \binits{G.J.D.}}:
\byear{2017},
\batitle{{Parameter optimization for surface flux transport models}}.
\bjtitle{Astron. Astrophys.}
\bvolume{607},
\bfpage{A76}.
\doiurl{https://doi.org/10.1051/0004-6361/201730689}.
\adsurl{2017A&A...607A..76W}.
\end{barticle}
\endbibitem

\bibitem[\protect\citeauthoryear{{Worden} and {Harvey}}{2000}]{worden2000}
\begin{barticle}
\bauthor{\bsnm{{Worden}}, \binits{J.}},
\bauthor{\bsnm{{Harvey}}, \binits{J.}}:
\byear{2000},
\batitle{{An Evolving Synoptic Magnetic Flux map and Implications for the
  Distribution of Photospheric Magnetic Flux}}.
\bjtitle{Solar Phys.}
\bvolume{195}(\bissue{2}),
\bfpage{247}.
\doiurl{https://doi.org/10.1023/A:1005272502885}.
\adsurl{2000SoPh..195..247W}.
\end{barticle}
\endbibitem

\bibitem[\protect\citeauthoryear{{Yeates}}{2014}]{yeates2014}
\begin{barticle}
\bauthor{\bsnm{{Yeates}}, \binits{A.R.}}:
\byear{2014},
\batitle{{Coronal Magnetic Field Evolution from 1996 to 2012: Continuous
  Non-potential Simulations}}.
\bjtitle{Solar Phys.}
\bvolume{289}(\bissue{2}),
\bfpage{631}.
\doiurl{https://doi.org/10.1007/s11207-013-0301-0}.
\adsurl{2014SoPh..289..631Y}.
\end{barticle}
\endbibitem

\bibitem[\protect\citeauthoryear{{Yeates}}{2016}]{yeatesdata}
\begin{botherref}
\oauthor{\bsnm{{Yeates}}, \binits{A.R.}}:
2016,
{Bipolar magnetic regions determined from NSO synoptic carrington maps}.
\doiurl{https://doi.org/10.7910/DVN/Y5CXM8}.
\url{https://doi.org/10.7910/DVN/Y5CXM8}.
\end{botherref}
\endbibitem

\bibitem[\protect\citeauthoryear{{Yeates}}{2020}]{yeatesdata2020}
\begin{botherref}
\oauthor{\bsnm{{Yeates}}, \binits{A.R.}}:
2020,
{Bipolar magnetic regions determined from HMI SHARPs data}.
\doiurl{https://doi.org/10.7910/DVN/1Z7YMT}.
\url{https://doi.org/10.7910/DVN/1Z7YMT}.
\end{botherref}
\endbibitem

\bibitem[\protect\citeauthoryear{{Yeates}, {Baker}, and {van
  Driel-Gesztelyi}}{2015}]{yeates2015}
\begin{barticle}
\bauthor{\bsnm{{Yeates}}, \binits{A.R.}},
\bauthor{\bsnm{{Baker}}, \binits{D.}},
\bauthor{\bsnm{{van Driel-Gesztelyi}}, \binits{L.}}:
\byear{2015},
\batitle{{Source of a Prominent Poleward Surge During Solar Cycle 24}}.
\bjtitle{Solar Phys.}
\bvolume{290}(\bissue{11}),
\bfpage{3189}.
\doiurl{https://doi.org/10.1007/s11207-015-0660-9}.
\adsurl{2015SoPh..290.3189Y}.
\end{barticle}
\endbibitem

\end{thebibliography}
%
% Without BibTeX 
% \begin{thebibliography}{}
% \bibitem[\protect\citeauthoryear{Author}{Year}]{key}
%   <bibliographical entry>
%
% \bibitem[\protect\citeauthoryear{}{}]{}
%   
%  
% \end{thebibliography}

\end{article} 
\end{document}